\documentclass[a4paper,11pt,preprint]{article}
\pdfoutput=1 
\usepackage{jheppub} 
\usepackage[T1]{fontenc} 

\usepackage{amsmath,amsfonts,amssymb,latexsym}
\usepackage{hhline}
\usepackage{graphicx}
\usepackage[arrow,matrix]{xy}
\usepackage{tikz}
\usepackage{tikz-cd}
\usetikzlibrary{positioning,arrows}
\usetikzlibrary{decorations.pathmorphing}
\usetikzlibrary{decorations.markings}

\usepackage{bm}
\usepackage{amssymb}
\usepackage{amsmath}
\usepackage{verbatim}
\usepackage{slashed}
\usepackage{mathrsfs}
\newcommand{\da}{{\dagger}}

\newcommand{\wh}{\widehat}
\newcommand{\p}{\partial}

\newcommand{\unit}{\mathbf{1}}

\newcommand{\ra}{\rightarrow}

\newcommand{\wt}{\widetilde}

\newcommand{\zz}{\mathbb{Z}}

\newcommand{\rr}{\mathbb{R}}

\newcommand{\mcc}{\mathcal{C}}

\newcommand{\mco}{\mathcal{O}}

\newcommand{\mcd}{\mathcal{D}}

\newcommand{\mcg}{\mathcal{G}}
\newcommand{\mcs}{\mathcal{S}}

\newcommand{\sca}{\mathscr{A}}

\newcommand{\vol}{\text{vol}\, }

\newcommand\be            {\begin{equation}}
\newcommand\ee            {\end{equation}}
\newcommand\ba            {\begin{aligned}}
\newcommand\ea            {\end{aligned}}

\definecolor{amethyst}{rgb}{0.6, 0.4, 0.8}

\title{Higher-form symmetries and spontaneous symmetry breaking}

\author[]{Ethan Lake}

\affiliation[]{Department of Physics, Massachusetts Institute of Technology, Cambridge, MA, USA}

\emailAdd{elake@mit.edu}

\abstract{We study various aspects of spontaneous symmetry breaking in theories that possess higher-form symmetries, which are symmetries whose charged objects have a dimension $p>0$. We first sketch a proof of a higher version of Goldstone's theorem, and then discuss how boundary conditions and gauge-fixing issues are dealt with in theories with spontaneously broken higher symmetries, focusing in particular on $p$-form $U(1)$ gauge theories. We then elaborate on a generalization of the Coleman-Mermin-Wagner theorem for higher-form symmetries, namely that in spacetime dimension $D$, continuous $p$-form symmetries can never be spontaneously broken if $p\geq D-2$. We also make a few comments on relations between higher symmetries and asymptotic symmetries in Abelian gauge theory.}

\begin{document} 
\maketitle
\flushbottom


\section{Introduction} 

Higher symmetries, also known as generalized global symmetries \cite{gaiotto2015generalized}, are symmetries whose charged objects have a dimension $p>0$: strings, membranes, volumes, and so on \cite{gaiotto2015generalized,kapustin2017higher,batista2005generalized}. Theories with higher symmetries are interesting for many reasons: they have been used to construct new models of topological phases \cite{yoshida2016topological}, to formulate an elegant description of relativistic hydrodynamics \cite{PhysRevD.95.096003,grozdanov2017generalised}, have been used in studies of elasticity theory and holography \cite{grozdanov2018generalised,hofman2017generalized}, and when they are discrete, their symmetry broken phases are topological field theories. This last point is especially interesting because it means that topological phases, which are often said to be ``beyond the symmetry-breaking paradigm'', can actually be understood within a symmetry-breaking framework \cite{gaiotto2015generalized}. In line with this, it would be nice to have concrete information about how higher symmetries can be broken, what types of phase transitions can exist between different higher symmetry-broken phases, and so on. In this paper, we take some small steps in this direction by studying various aspects of spontaneous symmetry breaking in theories with higher symmetries. 

We first offer a few general comments about $p$-form higher symmetries (see e.g. \cite{gaiotto2015generalized}), focusing on the symmetry group $U(1)$ for concreteness. The familiar case is $p=0$, which is a regular symmetry. The charged objects are exponentials $e^{i\phi}$, where $\phi$ is a compact scalar. Global symmetries correspond to the constant maps $\phi(x)=c$, which map all of spacetime $X$ to the same element $c\in \rr / 2\pi \zz$. More precisely, global symmetries are functions which are {\it locally} constant, that is, functions which are constant on each connected component of $X$. This means that the global symmetries are classified by a zeroth cohomology group, in our case $H^0(X;\rr/2\pi \zz)$. 
Such global symmetries give rise to conserved 1-form currents $J$ with $\p_\mu J^\mu = d^\da J=0$, and charge operators are constructed by 
\be Q(M_{D-1}) = \int_{M_{D-1}}\star J,\ee 
where $M_{D-1}$ is a codimension 1 manifold (e.g. all of space). 

Now for $p=1$. A theory with 1-dimensional objects charged under a $1$-form symmetry is a regular gauge theory. The charged objects are Wilson loops. They are constructed from a connection $A$, which is a connection on a principal $U(1)$-bundle over $X$ and tells us how to parallel transport zero-dimensional objects around the bundle. 
Since connections are completely determined by their holonomies, we can think of all the gauge-invariant information in $A$ as being captured by a map ${\rm hol}_A : \Omega X \ra U(1)$,
where $\Omega X$ is the loop space and the map assigns to each loop in $ X$ the holonomy of $A$ around the loop. 

Just as global symmetries in the $p=0$ case are the locally constant functions on $X$, global symmetries in the $p=1$ case are given by the locally constant functions $\lambda$ on $\Omega X$. These global symmetries are precisely the flat connections on $X$: they give rise to maps ${\rm hol}_\lambda$ which map all points in a given connected component of $\Omega X$, namely all the loops in a given homotopy class, to the same constant. They are the higher generalization of constant functions, and the analogue of the constant shift $\phi \mapsto \phi + c$ in the $p=0$ case is $A\mapsto A + \lambda$, where $d\lambda = 0$. When the $U(1)$ bundle is trivial, the global 1-form symmetries correspond to flat 1-forms on $X$, and as such are captured by a first cohomology group. 
These global symmetries give rise to conserved 2-form currents $F$ with $d^\da F=0$, and charge operators are constructed as $Q(M_{D-2}) = \int_{M_{D-2}}\star F$, where $M_{D-2}$ is a codimension 2 manifold. 

For $p=2$ the charged objects are two-dimensional Wilson surfaces, which are constructed from a connection $B$ on a principal $U(1)$ bundle over $\Omega X$. In the $p=1$ case we have a connection $A$ which allows us to parallel transport points along loops, while in the $p=2$ case we have $B$, which allows us to parallel transport strings along surfaces. 
Just as we can define a connection as a function ${\rm hol}_A$ from loops in $X$ to $U(1)$, we may think of $B$ as providing a way to map surfaces in $X$ to $U(1)$, where the map ${\rm hol}_B$ is given by integrating $B$ along a given surface. Global symmetries in this case are the analogues of constants for 2-forms: they are flat connections on $\Omega X$, or in the case when the bundle is trivial, flat 2-forms on $X$, which are captured by a second cohomology group. 
Such symmetries lead to conserved 3-form currents $K$, with charge operators $Q(M_{D-3}) = \int_{M_{D-3}}\star K$. 

The pattern for $p>2$ is the same: global $U(1)$ $p$-form symmetries arise when we consider a $p$-form field $A$ which specifies a collection of maps (the Wilson loop / Wilson brane operators) from $p$-dimensional submanifolds of $X$ to $U(1)$. The global symmetry acts on $A$ by shifting it by the generalized notion of a constant function. That is, the global symmetry shifts $A\mapsto A+\lambda$, where $\lambda$ is a flat $p$-connection. For trivial bundles, $\lambda$ is a closed $p$-form on $X$, and so the nontrivial higher $p$-form symmetries are parametrized by a $p$-th cohomology group. 

Translating $A$ by a flat $p$-form is a global symmetry (rather than a local one), since it is impossible to do this translation locally. For example, if $X = S^1\times \rr^d$, shifting a 1-form $A$ by a flat 1-form $\lambda$ that has nontrivial holonomy around the $S^1$ necessarily changes $A$ along the whole of $\rr^d$. Thus these are {\it not} local symmetries, even though they act directly on $A$, which is a gauge field\footnote{Importantly, these are also {\it not} a subset of topologically nontrivial (or ``large'') gauge transformations (gauge transformations that are not in the identity component of the group of gauge transformations), since they in general cannot be written as $\lambda = igdg^{-1}$ for a well-defined function $g$.}. 

A strong and simplifying restriction on theories with higher-form symmetries comes from the fact that if $p>0$, the symmetry group must be Abelian \cite{henneaux1986p,gaiotto2015generalized}. This is because by current conservation $d\star J=0$, the charge operators $Q(M) \sim \int_M \star J$ are topological, meaning that as long as they do not intersect charged operators their values only depend on the homology class of $M$. If the codimension of the charge operators is greater than 1 (i.e. if $p\geq1$), then the operator $Q_1Q_2$ can always be smoothly deformed into $Q_2Q_1$, implying the charge algebra is Abelian (when $p=0$ the $Q_i$ are supported on codimension 1 submanifolds, and there is ``no room'' for them to be moved around one another). 
We will be mostly interested in continuous higher symmetries for generic $p$, and so we will restrict our attention to theories with a $U(1)$ $p$-form symmetry in what follows.  

The structure of the remainder of this paper is as follows: we sketch a proof of a higher version of Goldstone's theorem in Section \ref{goldstone}, and show how symmetry-broken phases of theories with continuous higher symmetries can be understood as $p$-form $U(1)$ gauge theories, where the charged objects are higher-dimensional analogues of Wilson loops. 
In \ref{charge_ops} we discuss these $p$-form theories in more detail, focusing on the properties of charge operators. We then give a treatment of the important issue of boundary conditions in Sections \ref{boundary_conditions}, write down Ward identities arising from higher symmetries in Section \ref{ward_ids}, and discuss gauge-fixing issues in Section \ref{gauge_fixing}.
In Section \ref{asymp_syms}, we briefly discuss higher-form SSB on manifolds with trivial cohomology (where the order parameters are Wilson operators that end on the boundary or go off to infinity), and offer some comments on a potential relation between these symmetries and the asymptotic symmetries of $U(1)$ gauge theories.

Section \ref{CMW} is devoted to proving an analogue of the Coleman-Mermim-Wagner theorem for higher symmetries, namely that continuous $p$-form symmetries in $D$ spacetime dimensions can never be broken if $p\geq D-2$ (for discrete symmetries, this is changed to $p\geq D-1$). This theorem was stated in \cite{gaiotto2015generalized}; here we elaborate on the details and examine the theorem a bit more closely. We give proofs both for ``compact'' theories with magnetic matter (lattice theories), and theories without them. In the former case, our proof is a simple generalization of Polyakov's classic argument \cite{polyakov1977quark} showing confinement for $U(1)$ lattice gauge theory in $D=3$ (which is equivalent to the statement that $p=1$ form symmetries cannot be broken in $D=3$) to general dimensions and general form fields. For theories without magnetic matter the flavor of the proof is different, and is based on computing expectation values of generalized Wilson loop operators (we are aware that similar results have been obtained by S. Grozdanov \cite{grozdanov2018talk}). 

We now summarize our notational conventions. $X$ will denote spacetime, which we assume throughout to be orientable and to have torsion-free homology for simplicity. 
$\Omega^p(X)$ will denote all $p$-forms on $X$, and $\Omega^p_\zz(X)$ will denote those with integral periods around closed $p$-submanifolds. $C_p(X)$ will denote the $p$-chains (submanifolds) on $X$, and $C^p(X)$ the $p$-cochains. $Z_p(X)$ will be used for closed $p$-chains, and $Z^p(X)$ for closed $p$-forms / $p$-cochains. Unless otherwise indicated, all coefficients will be taken in $\rr$. 
We will write the codifferential as $d^\da$, which when acting on $p$-forms in Euclidean signature is $d^\da = (-1)^{Dp + D +1}\star d \star,$ with $\star$ the Hodge dual and $D$ the dimension of the ambient manifold. Our conventions regarding exterior calculus and some background on various mathematical tools can be found in Appendix \ref{forms}.

Finally, a note on terminology: we will use the term ``trivial gauge transformation'' to describe a transformation that shifts a gauge potential $A$ by an exact form, and does not affect the physical data of the theory (like the boundary conditions). These transformations are always gauged. This is in contrast to gauge transformations which shift $A$ by an exact form, but which act nontrivially on the physical data of the theory. A necessary requirement is that such transformations must be ``large'', but this is not sufficient: sufficiency is theory-dependent, and in particular depends on the choice of boundary conditions. 

\section{A higher Goldstone theorem and $p$-form theories} \label{gs_and_pforms} 

\subsection{Goldstone's theorem for $p$-form symmetries} \label{goldstone}

In this section, we sketch a simple proof of Goldstone's theorem for $p$-form symmetries. We will see that symmetry-broken phases of theories with $p$-form symmetries are deconfined phases of tensionless $p$-branes, where the Goldstones (which will be $p$-form gauge fields) arise as the massless modes associated with transverse diffeomorphisms of the branes. 

There are many proofs of Goldstone's theorem for the $p=0$ case; here we present a version of one \cite{guralnik1967broken} which is most amenable to generalization to the arbitrary $p$ case\footnote{One common way to prove Goldstone's theorem with minimal assumptions is to study the theory in the presence of a small symmetry breaking field; it is not obvious how to perform the same calculation in the $p>0$ case since the order parameters one needs to add to the action are not local (at least, not without going to loop space).}. 

For each $p$-form symmetry we have a conserved $(p+1)$-form current $J$, from which we construct the charge operator 
\be Q(M) = \int_M \star J,\ee
where the integral is taken over a $(D-p-1)$-manifold $M\subset \Sigma$, with $\Sigma$ a $(D-1)$-manifold which we think of as a constant time slice, or any other codimension 1 Cauchy surface. Note that when we write $\star$, we mean the Hodge dual taken with respect to the full spacetime $X$, and not with respect to $\Sigma$. 

Using Poincare duality, we can rewrite the charge operator as 
\be Q(M) = \int_\Sigma \star J\wedge \wh M,\ee 
where $\wh M$ is a $p$-form which is the Poincare dual (with respect to $\Sigma$) of $M$. 

Some brief comments on Poincare duality: in general, if $M\subset \Sigma$ is any codimension $p$ submanifold of $\Sigma$, its Poincare dual $\wh M\in \Omega^p(\Sigma)$ is a normalized bump function that is oriented transversely to $M$ and has compact support on a tubular neighborhood of $M$ in $\Sigma$ (see Appendix \ref{forms} for more detail). For example, if $\Sigma=\rr^2$ and $M$ is the $y$-axis, then $\wh M$ is the 1-form $\rho(x)dx$, with $\rho(x)$ a bump function centered at $x=0$. 
The dual of the exterior derivative is the boundary operator, so that $\p M = d(\wh M)$.

Now, let $\mco$ be an operator which is charged under a global $p$-form symmetry, like the Wilson operator $\mco = W_C = \exp(i\int_CA)$, which is defined for some $p$-manifold $C$ and $p$-form $A$. Consider a scenario in which SSB occurs, and define $\mcc$ by 
\be \mcc = \langle 0|[Q(M),\mco] |0 \rangle\ee
for some reference symmetry-breaking vacuum state $|0\rangle$ in which $\langle0| \mco |0\rangle \neq 0$. Since $\mco$ is charged under the $p$-form symmetry we can choose $M$ so that $\mcc\neq0$: for example if $\mco = W_C$, we can choose $M$ so that $M$ intersects $C$ once\footnote{This is possible since $C$ is not a boundary in $C_p(\Sigma;\zz)$ by the assumption that $W_C$ is charged under the symmetry generated by $Q(M)$ (more on this later).}. 

We can now simply proceed as in the $p=0$ case, since $\star J \wedge \wh M$ is a 0-form on $\Sigma$. Inserting a complete set of states,
\be \ba \mcc & = \sum_n\int_\Sigma d^{D-1}x\, \Big(\langle 0| (\star J \wedge \wh M)(x)|n\rangle \langle n | \mco |0\rangle - \langle 0| \mco |n\rangle \langle n |  (\star J \wedge \wh M)(x) |0\rangle\Big).\ea\ee
Now we 
do the integral over $x$:
\be\ba \label{final_C_form} \mcc & = \sum_n(2\pi)^{D-1} \delta^{D-1}(\mathbf{p})\Big(\langle 0| (\star J \wedge \wh M)(0)|n\rangle \langle n | \mco |0\rangle e^{-i\omega_nt} - \langle 0| \mco |n\rangle \langle n |  (\star J \wedge \wh M)(0) |0\rangle e^{i\omega_nt}\Big),\ea\ee
where $t=x^0$ is the direction of the unit normal to $\Sigma$. 

We will now show that the RHS of \eqref{final_C_form} is actually independent of $t$. Differentiating $\mcc$ with respect to $t$,
\be \p_0 \mcc = \int_{M} \langle 0 | [\p_0 \star J,\mco] | 0 \rangle,\ee
where we have used Poincare duality to trade the wedge product with $\wh M$ for an integral over $M$. 

By current conservation $d^\da J=0$, we have $\p_0 J^{0\mu_1\dots \mu_p} = -\p_jJ^{j\mu_1\dots\mu_p}$, which lets us write 
\be \label{po_mcc} \p_0\mcc = -  \int_{\p M \subset \p\Sigma} \langle 0 | [\star_\Sigma J,\mco] | 0 \rangle.\ee
Here, $\star_\Sigma$ is the Hodge dual restricted to the spatial slice $\Sigma$. Since $\Sigma$ is a $(D-1)$-manifold, $\star_\Sigma J$ is a $(D-1)-(p+1)=(D-p-2)$-form, matching with the fact that $\p M$ is a $(D-p-2)$-dimensional submanifold. 
 
The expression on the RHS of \eqref{po_mcc} can be made to vanish if 
\be \label{supp_cond} {\rm Supp}(\mco) \cap \p M = 0,\ee
where ${\rm Supp}(\mco)$ denotes the support of $\mco$. For any generic choice of $\mco$, we can always choose an $M$ such that \eqref{supp_cond} is satisfied and $\mcc \neq 0$\footnote{This is just a consequence of counting dimensions: ${\rm Supp}(\mco)$ is a $p$-dimensional submanifold of $X$. In the case that ${\rm Supp}(\mco)\subset \Sigma$, then ${\rm Supp}(\mco)$ is a $(p-1)$-dimensional submanifold of $\p \Sigma$, while $\p M$ is a $(D-p-2)$ submanifold of $\p \Sigma$. Therefore $\dim {\rm Supp}(\mco) + \dim \p M < \dim \p \Sigma$ and so \eqref{supp_cond} is generically satisfied (if ${\rm Supp}(\mco)\not\subset\Sigma$, the inequality is stronger).}. 

Therefore, 
we then see that $\p_0\mcc=0$, and yet the RHS of \eqref{final_C_form} carries time dependence in the exponentials. Hence if $\mcc$ is non-zero, then we must have states $n$ such that $\omega_n =0$ when $\mathbf{p}=0$, i.e. the spectrum must be gapless at zero momentum. This tells us that we should expect Goldstone modes to appear in the theory. 

Given a theory with a spontaneously broken $p$-form symmetry, we want to identify the Goldstone modes. Because they are created by $(p+1)$-form conserved currents, they will be $p$-form fields. They are precisely the fields that shift linearly under the symmetry action. 
For $p=0$, this is the familiar shift $\phi \mapsto \phi + c$, for $c$ a constant. 
As mentioned earlier, the analogue of this for $p>0$ is $A\mapsto A + \lambda$, for $\lambda$ a flat $p$-form which is not a trivial (local) gauge transformation (technically, a flat connection in the case of a nontrivial bundle). Thus, the Goldstones are $p$-form gauge fields.

Given that $p$-form symmetries are always Abelian for $p>0$, if Goldstones are the only massless fields, essentially the only effective IR action we can write down is the kinetic term for the Goldstones, namely the generalized Maxwell action
\be \label{pform_action} S = -\frac{1}{2g^2}\int_X F\wedge \star F,\ee
with $F=dA$\footnote{Depending on $D$ and $p$, we may also add a Chern-Simons term or a $\theta$ term. The former is not invariant under the higher symmetry $A\mapsto A + \lambda$ since the boundary term $\int_{\p X} \lambda \wedge A$ is generically non-zero, and we will ignore the latter for simplicity since we are mostly focused on general choices of $p,D$, for which a $\theta$ term usually does not exist.}. 

Note that the action \eqref{pform_action} actually has two symmetries, related by electromagnetic duality: the $p$-form symmetry with conservation law $d^\da F=0$ which we call the ``electric'' symmetry, and a $(D-p-2)$-form ``magnetic'' symmetry arising from the conservation law $d^\da\star F=0$ \cite{gaiotto2015generalized}. The electric symmetry shifts $A\mapsto A + \lambda$ for $\lambda \in Z^p(X)$ while the magnetic one shifts $\wt A \mapsto \wt A + \wt \lambda$ for $\wt \lambda \in \Omega^{D-p-2}(X)$, where $\star F = d\wt A$. The electric symmetry is the higher form version of the momentum symmetry of the compact scalar, while the magnetic symmetry is the higher generalization of the winding number symmetry. These currents are conserved only in the absence of dynamical matter: electrically charged matter explicitly breaks the conservation equation $d^\da F=0$, and translating $A$ by a nontrivial flat connection is no longer a global symmetry as it leads to inconsistent boundary conditions for the matter fields. Likewise, the presence of magnetically charged matter explicitly breaks $d^\da \star F=0$. In what follows we will work in the context of pure gauge theory without either type of matter, unless specified otherwise. 

Note that when we set $p=1$, the symmetry-broken phase is described by regular Maxwell electromagnetism. This means that in the language of higher symmetries, we may interpret the photon as a Goldstone boson of a spontaneously broken 1-form symmetry (provided that we are in a scenario in which a symmetry-broken phase is allowed). This interpretation \cite{gaiotto2015generalized,kovner1991photon,strominger2017lectures} was actually noted very early on by Polyakov \cite{polyakov1980gauge}, who viewed the photon as a Goldstone boson realized as a scalar field on the loop space $\Omega X$.

\subsection{Canonical commutators and charge operators} \label{charge_ops}

Now we discuss some more general aspects of higher symmetries in $p$-form gauge theories, which we will need in subsequent sections (see also parts of Section 3 of \cite{gaiotto2015generalized}). 


Consider a codimension 1 submanifold $\Sigma \subset X$ of spacetime; $\Sigma$ may be a spatial slice or any other Cauchy surface. The variational 1-form associated to the action \eqref{pform_action} is 
\be \theta_\Sigma = -\int_\Sigma\frac{1}{g^2}\star F\wedge \delta A,\ee 
where $\delta$ is the variational exterior derivative. This means that the symplectic form $\Omega = \delta \theta_\Sigma$ is 
\be \label{symplectic_form} \Omega_\Sigma = -\int_\Sigma\delta\left(\frac{1}{g^2}\star F\right)\wedge\delta A.\ee
$\Omega_\Sigma$ has zero modes as it stands, an issue which will be fixed when we discuss boundary conditions and gauge fixing. 

The coordinates in the phase space are thus the gauge field $A$ and the electric field $\frac{1}{g^2}\star F$. 
The form of $\Omega_\Sigma$ leads to the commutator 
\be [(\star F)_{\mu_1\dots\mu_{D-p-1}}(x),A_{\nu_1\dots\nu_p}(y)] = -ig^2\varepsilon_{0\mu_1\dots\mu_{D-p-1}\nu_1\dots\nu_p}\delta(x-y),\ee
where $0$ is the direction normal to $\Sigma$. 
The commutation relation here means that explicitly, 
\be \label{F_as_deltaA} \frac{1}{g^2} (\star F)(x) = -i \frac{1}{p!(D-p-1)!} \frac{\delta}{\delta A_{\mu_1\dots\mu_p}(x)}\varepsilon_{0\mu_1\dots\mu_p\mu_{p+1}\dots\mu_{D-1}}dx^{\mu_{p+1}}\wedge\dots\wedge dx^{\mu_{D-1}},\ee
where the $1/p!$ factor is needed to absorb the over-counting coming from the anti-symmetrization of the indices in $A$. 

This means that 
the charge operator measuring the flux passing through a $(D-p-1)$-manfiold $M\subset \Sigma$, namely 
\be Q(M) = \frac{1}{g^2}\int_M \star F,\ee
is the operator which acts as the translation operator on gauge fields.
To see why, we use Poincare duality to write $\int_M\star F = \int_\Sigma \star F \wedge \wh M$\footnote{Again, $\wh M$ is the Poincare dual of $M\subset \Sigma$ with respect to $\Sigma$, not with respect to the full spacetime $X$. Thus $\wh M$ is a $(D-1)-(D-p-1)=p$-form.}, and we see that for $\alpha \in \rr$,
\be e^{i \alpha Q(M)}Ae^{-i \alpha Q(M)}= \exp\left(\frac{i\alpha}{g^2}\int_{\Sigma}\star F\wedge \wh M\right) A \exp\left(-\frac{i\alpha}{g^2}\int_{\Sigma}\star F\wedge \wh M\right)= A + \alpha\wh M.\ee

The Wilson operators 
\be W_C = \exp\left(i\int_C A\right)\ee 
for $C\in C_p(\Sigma;\zz)$
are charged under $Q(M)$ if the (signed) intersection number of $M$ and $C$ is nonzero. Explicitly, we can use Poincare duality to see that 
\be \label{expQonW} e^{i\alpha Q(M)}W_Ce^{-i\alpha Q(M)} = e^{i\alpha I(M,C)} W_C,\ee
where $I(M,C)\in \zz$ is the signed intersection number,
\be I(M,C)= \int_\Sigma \wh M \wedge \wh C = \int_{M\cap C}1, \ee 
which is simply the charge of $W_C$ under $Q(M)$. Since we take both $M$ and $C$ to define $\zz$-valued chains, $I(M,C)\in\zz$ is always an integer. Note that $I(M,C)$ can only be non-zero if $M$ and $C$ are not boundaries in $C_{D-p-1}(\Sigma;\zz)$ and $C_p(\Sigma;\zz)$, respectively. 

While we usually take $M\subset \Sigma$, we can also tilt $M$ out of $\Sigma$. The charge operator is still the same, and we still have a Ward identity relation like \eqref{expQonW}, except that in this case the intersection number $I(M,C)$ is replaced with the linking number. In such cases, the action of $Q(M)$ can be treated by using current conservation to deform $M$ to lie within $\Sigma$, which can be done as long as $M$ is not deformed through any charged operators. 

The charge $\wt Q$ for the dual $(D-p-2)$-form magnetic symmetry is obtained by integrating the current $\frac{1}{2\pi}F$ over a $(p+1)$-manifold $N\subset \Sigma$:
\be \wt Q(N) = \frac{1}{2\pi} \int_N F = \frac{1}{2\pi} \int_\Sigma F \wedge \wh N,\ee
which shifts the dual gauge field $\wt A\in \Omega^{D-p-2}(X)$ by $\wt A \mapsto \wt A + \wh N$, where $\star F = d\wt A$ and $\wh N$ is a $(D-p-2)$-form. 

The Wilson operators $W_C$ are well-defined only if they are gauge-invariant, and they are only invariant under a gauge transformation $A\mapsto A + d\varepsilon$ if $\int_{\p C} \varepsilon=0$. Thus, $W_C$ is gauge invariant if $\p C=0$, but it is also gauge-invariant if $\p C \subset \p\Sigma$, provided that boundary conditions on $\p \Sigma$ are fixed so that gauge transformations are required to vanish at $\p \Sigma$. Assuming that this is the case, $W_C$ is gauge-invariant if $C$ is closed modulo $\p \Sigma$. This is formalized by saying that $C\in Z_q(\Sigma,\p \Sigma;\zz)$, where
$Z_q(X,\p X;\zz)$ denotes all the $q$-chains on $X$ which are relatively closed, i.e. which are closed modulo $\p X$\footnote{Relative cohomology groups $H^k(X,\p X)$ consist of closed $k$-forms that vanish on $\p X$ modulo exterior derivatives of $(k-1)$-forms that vanish on $\p X$, while relative homology groups $H_k(X,\p X)$ contain $k$-submanifolds whose boundaries are contained within $\p X$, modulo $k$-manifolds which are boundaries modulo $\p X$. The definition for relative (co)chains and (co)cycles is analogous.}.

\begin{figure}
\centering
\includegraphics{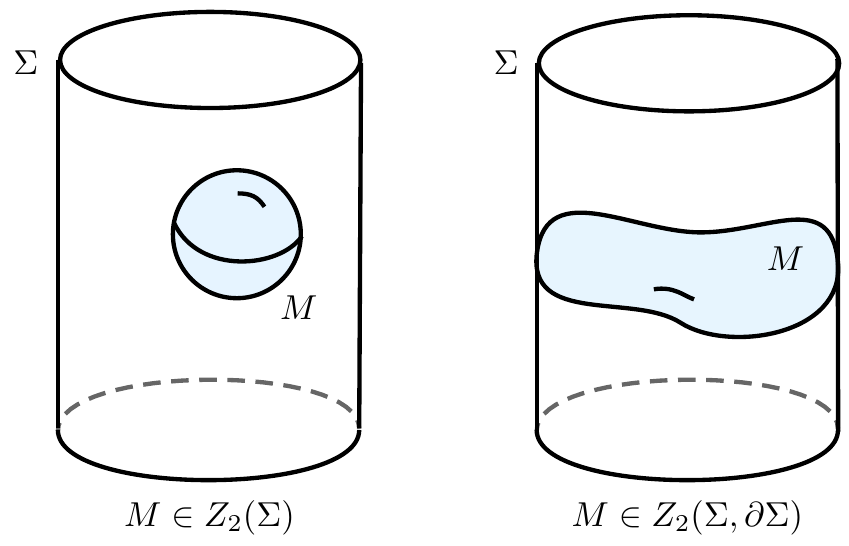}
\caption{\label{boundaries_fig} An example of the support of two charge operators $Q(M)$ inside a tube $\Sigma$. On the left the 2-manifold $M$ is closed with $\p M=0$, while on the right it is relatively closed: $\p M\neq0$, but $\p M \subset \p \Sigma$. The left $M$ is a boundary, and thus the associated $Q(M)$ generates a trivial gauge transformation. }
\end{figure}

$Q(M)$ with $M\subset \Sigma$ only generates a global symmetry of the action if it translates $A$ by a flat form, i.e. if $d\wh M =0$, where again the Poincare dual is taken on $\Sigma$. Therefore, we can identify operators $Q(M)$ that generate global symmetries by examining $Z^p(\Sigma)$, the closed $p$-forms on $X$.  
Poincare duality on $\Sigma$ produces an isomorphism 
\be \label{pduality_on_sigma} Z^p(\Sigma) \cong Z_{D-p-1}(\Sigma,\p \Sigma),\ee 
and so $Q(M)$ only generates a global higher symmetry if $M \in Z_{D-p-1}(\Sigma,\p \Sigma)$ is a relative cycle. In other words, $Q(M)$ only generates a symmetry if $\p M \subset \p \Sigma$. If we had $\p M \not\subset\p \Sigma$, one could integrate $A$ along a closed submanifold linking $M$ only once, which would give rise to nontrivial holonomies for $A$ along contractible loops, leading to a non-flat transformation (which is not a symmetry of the action). An illustration of the difference between closed $p$-chains $Z_p(\Sigma)$ and relatively closed $p$-chains $Z_p(\Sigma,\p \Sigma)$ is shown in Figure \ref{boundaries_fig}. 

In the case where $\wh M = d\varepsilon $ is exact, $Q(M)$ is precisely the Hamiltonian generator of gauge transformations which one obtains from the symplectic form $\Omega_\Sigma$, since it translates $A\mapsto A + d\varepsilon$. From the duality \eqref{pduality_on_sigma}, the condition that $\wh M$ is exact means that $M$ is a relative boundary in $C_{D-p-1}(\Sigma,\p \Sigma)$. That is, it implies the existence of a $(D-p)$ manifold $B$ such that 
\be \p B \setminus (\p B \cap \p \Sigma) = M.\ee 

If $\wh M$ is both exact and has no support on $\p \Sigma$ so that $M=\p B$ is an absolute boundary, then $M$ defines a trivial class in $H_{D-p-1}(\Sigma;\zz)$ as well as in $H_{D-p-1}(\Sigma,\p \Sigma;\zz)$. In this case $Q(M)$ vanishes on-shell. Indeed, writing $\wh M = d \varepsilon$ with $\varepsilon|_{\p \Sigma}=0$ as a consequence of $M\cap \p \Sigma =0$, we see that 
\be Q(M)=\frac{1}{g^2}\int_M\star F = \frac{1}{g^2}\int_\Sigma \star F \wedge d\varepsilon = (-1)^{D-p}\frac{1}{g^2}\int_\Sigma d\star F \wedge \varepsilon = 0.\ee 
$Q(M)$ vanishes in this case because it generates a local (trivial) gauge transformation. 
Of course, the Wilson operators $W_C$ for all $C\in Z_p(\Sigma,\p \Sigma;\zz)$ are invariant under such a gauge transformation, since the intersection number $I(M,C)=0$ for such $M$. 

In contrast, if $\wh M$ is exact but $\p M \cap \p \Sigma \neq 0$, then $M$ then is a boundary in $C_{D-p-1}(\Sigma,\p \Sigma;\zz)$, but not in $C_{D-p-1}(\Sigma;\zz)$. This means that $\wh M = d\varepsilon$ is exact but has support at $\p \Sigma$, where boundary conditions on $A$ may be defined. In this case, the charge operator $Q(M)$ does not necessarily generate a local gauge symmetry, and can still in fact generate a global symmetry, depending on the choice of boundary conditions. We will comment more on this when we discuss boundary conditions and asymptotic symmetries in the following sections. 

Finally, if $\wh M$ is not exact, then it shifts $A$ by a flat connection which is not a gauge transformation, thereby generating a global symmetry. 
The Poincare dual version of this statement is that integrating the electric flux on a closed manifold that is a boundary generates a gauge transformation, while doing the integral over a manifold that is closed but not a boundary generates a global higher-form symmetry. 

The fact that the symmetry generators $Q(M)$ are given by integrals over manifolds $M$ that are not boundaries means that one cannot continuously connect a vacuum state with one set of expectation values for Wilson loops to a vacuum state with a different set of expectation values. Thus, the expectation value of the phase of the $W_C$ Wilson operators for each $[C]$ provide labels of different superselection sectors in the theory, where $[C]$ is the equivalence class of all $C'$ which can be deformed to $C$ by a homotopy which preserves $\p C'=\p C$. These superselection sectors cannot be connected with local operators and constitute the different symmetry-broken vacua. 

\begin{figure}
\centering
\includegraphics{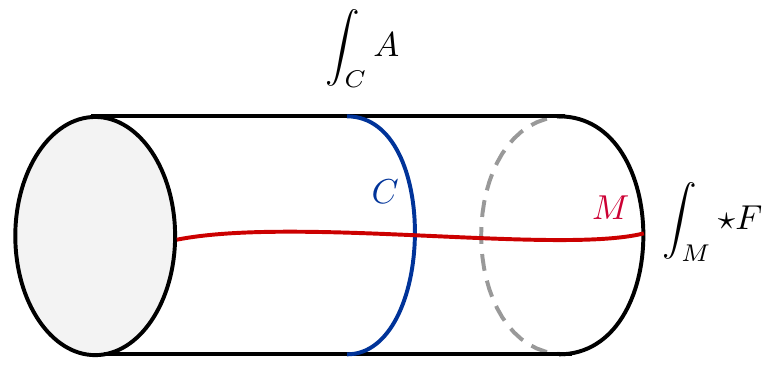}
\caption{\label{tube} A section of a tubular spatial slice of the form $S^1\times \rr$. Integrating $\star F$ along the manifold $M$ (red line) measures the electric flux around the $S^1$. Acting with $Q(M)$ transforms a Wilson loop wrapping around the $S^1$, which is given by the integral of $A$ along $C$ (blue line).}
\end{figure}

For example of a typical setup, consider regular Maxwell theory ($p=1$) on a spacetime of the form $(S^1\times \rr)\times \rr$, where space is the tube $S^1\times \rr$, drawn in Figure \ref{tube}. Wilson loops which wrap the $S^1$ are charged under a 1-form symmetry which translates $A_\tau \mapsto A_\tau + (\gamma / \beta)d\tau$. Here $\tau$ is the direction of the $S^1$, which has a circumference of $\beta$, and $\gamma \in \rr/2\pi \zz$. The form $( \gamma / \beta)d\tau$ has a holonomy of $\gamma$ around the $S^1$, and so the Wilson operator shifts as $W_C \mapsto e^{2\pi i \gamma}W_C$ under the symmetry, which is generated by 
\be \exp(iQ(M)) = \exp\left(i \frac{\gamma}{g^2}\int_{C}(\star F)_xdx \right),\ee
where $C$ is the axis along the spatial $\rr$. Changing the phase of $\langle W_C\rangle$ in a given state can only be done by integrating $\star F$ along the (infinite) length of the cylinder, and so the phase of $\langle W_C\rangle$ can be used to label superselection sectors.

\section{Boundary conditions and gauge-fixing} \label{bconds_and_gfing}

\subsection{Boundary conditions \label{boundary_conditions}} 

We now turn our attention the sometimes subtle issue of boundary conditions, which are very important for studying phases with symmetry breaking. From our perspective, SSB occurs when physics inside a system develops a strong dependence on the choice of boundary conditions.   Indeed, in the absence of external fields, SSB can only happen on spacetimes which are not closed, as boundary conditions are required to specify how the symmetry breaking occurs. 

When working on an unbounded spacetime $X$, specifying boundary conditions usually means specifying asymptotic falloff conditions at infinity. However, much of our analysis of symmetry breaking on unbounded $X$ can be done by imposing an IR cutoff by taking $X$ to be large but finite, with a hard boundary $\p X$ on which we impose boundary conditions. This will be our approach in most of what follows. 

First, some notational preliminaries. 
There are two types of boundary conditions that we will often want to impose on fields. The first is Dirichlet boundary conditions, where we set the boundary components of the field to zero. We will use the notation 
\be \Omega^p_D(X) = \{ \alpha \in \Omega^p(X) \; : \; \alpha|_{\p X} =0\}\ee
to denote Dirichlet $p$-forms\footnote{To be precise, by $\alpha|_{\p X}=0$ we mean $\iota^*_\p\alpha=0$, where $i_\p : \p X \ra X$ is the inclusion.}. 

The second is Neumann (or ``tangential'') boundary conditions, where we set the components of the field normal to the boundary to zero. We will use the notation 
\be \Omega^p_N(X) = \{ \alpha \in \Omega^p(X) \; : \; (\star\alpha)|_{\p X} =0\}\ee
to denote Neumann $p$-forms, so that if $\alpha \in \Omega^p_N(X)$, $\alpha$ vanishes when contracted with any vector normal to the boundary. Note that the two boundary conditions are dual under $\star$, in that if $\alpha \in \Omega^p_D(X)$ then $\star \alpha \in \Omega^p_N(X)$, and vice versa. 
These boundary conditions are natural because then for $\langle \alpha,\beta\rangle = \int_X \alpha \wedge \star \beta$ the inner product on $\Omega^p(X)$, we have $\langle d \alpha,\beta\rangle = \langle \alpha, d^\da\beta\rangle$ (implying the Laplacian $\Delta = (d+d^\da)^2$ is self-adjoint) only when $\alpha,\beta$ satisfy either Neumann or Dirichlet boundary conditions.

For the $p$-form theories, the on-shell variation of the action is 
\be \delta S = -\frac{1}{g^2}\int_{\p X} \star F \wedge \delta A = -\frac{1}{g^2} \int_{\p X} d^{D-1}x^\mu_\perp\; F_{\mu\nu_1\dots\nu_p}A^{\nu_1\dots\nu_p}.\ee
This must vanish, and we have a few ways of making it do so.
One option is to let the fields be unconstrained on $\p X$, but to add boundary degrees of freedom that cancel $\delta S$. This option is less well-suited to a discussion of symmetry breaking, and we will not make use of it here. 

There are two more options, which are electromagnetically dual to one another. 
One option is to set boundary conditions by requiring that $A|_{\p X}$ be fixed to some specified vacuum configuration. This means that for variations preserving the boundary conditions, $(\delta A)_{\nu_1\dots\nu_p}\neq 0$ only if one of the $\nu_1$ is the direction normal to the boundary; this ensures that $\delta S=0$ by the antisymmetry of $F$. 
The vacuum configurations have $F|_{\p X}=0$, so that $F\in \Omega_D^{p+1}(X)$. 
Therefore, different vacua (which will only actually be distinct if SSB is allowed) are labeled by flat $p$-forms on $\p X$. 
In locations where $\p X$ has a timelike component, this corresponds to setting both the electric field parallel to the boundary and the magnetic field normal to the boundary to be zero --- these are the boundary conditions (``electric'' boundary conditions) one has between vacuum and a superconductor. 

With these boundary conditions, Wilson operators $W_C$ that end on the boundary (i.e. those with $\p C \subset \p X$) are gauge-invariant, since gauge transformations are not allowed to change the boundary conditions on $A|_{\p X}$. Thus just as at the boundary between vacuum and a supercondcutor, electric field lines (and their $p$-form generalization, Wilson branes) are allowed to end on the boundary, while conversely magnetic field lines (t'Hooft branes) are not. 

The second option for boundary conditions is to set $\delta S=0$ by requiring that $F\in \Omega^{p+1}_N(X)$ so that $\star F|_{\p X}=0$. 
Since this fixes $\star F|_{\p X}$, we must allow $A|_{\p X}$ to be unconstrained. This sets both the electric field normal to the boundary and the magnetic field parallel to the boundary to be zero --- these are the boundary conditions (``magnetic'' boundary conditions) between a monopole condensate and the vacuum. 
With these boundary conditions, Wilson operators that end on the boundary are {\it not} gauge-invariant, since gauge transformations needn't preserve $A|_{\p X}$, which is unconstrained. That Wilson operators are not allowed to end on the boundary with this set of boundary conditions is the electromagnetic dual of the ($p$-form generalization of the) statement that superconductors expel magnetic flux. 

 The magnetic boundary conditions still allow degenerate vacua. Let $\star F = d\wt A$, where $\wt A\in \Omega^{D-p-2}(X)$ is the magnetic dual potential. Since $\star F|_{\p X}=0$, we require that $\wt A$ be a flat $(D-p-2)$-form on $\p X$, and so the different vacua are given by specifying $\wt A|_{\p X} \in Z^{D-p-2}(\p X)$. 

The two boundary condition schemes are related to the two higher-form symmetries of the theory: the electric boundary conditions are related to the $p$-form symmetry of $A$, and the magnetic ones are related to the $(D-p-2)$-form magnetic symmetry of $\wt A$. Note that only when 
\be p=\frac{D-2}{2}\ee 
(e.g. regular electrodynamics or $D=4$ or the compact scalar in $D=2$) do we have a self-duality where both of these symmetries parametrized by the same forms (also, only when $p=(D-2)/2$ is the theory conformally invariant). 

When we are interested in discussing SSB for the $p$-form symmetry, we must choose electric boundary conditions (when we refer to ``SSB'' without an epithetical adjective, we mean SSB of the $p$-form electric symmetry). We need to make this choice since electric boundary conditions are the ones which let us specify the values of the order parameters $W_C$ on initial and final states, and are the ones for which the $W_C$ operators are allowed to terminate on the boundary $\p X$ in a gauge-invariant way. If we chose magnetic boundary conditions instead then $A|_{\p X}$ would be free, and we would be unable to use boundary conditions to select out different symmetry-broken ground states for the $p$-form symmetry. Likewise, discussing magnetic SSB would require choosing magnetic boundary conditions. 

In the symmetry-broken phase of the $p$-form theory with electric boundary conditions, the different symmetry-breaking states are distinguished by expectation values of the $W_C$ operators on the boundary. We can prepare a state on an initial spatial slice $\Sigma$ by translating the fields in some reference state $|0\rangle$ with the help of the charge operators. We write these states as
\be \label{states} |\alpha,M\rangle = e^{i\alpha Q(M)} | 0\rangle,\ee
where $M\subset Z_p(\Sigma,\p\Sigma;\zz)$. The notation $|\alpha, M\rangle$ indicates a state in which, assuming symmetry-breaking, the phase of $\langle W_C\rangle$ with $C$ contained in a spatial slice is $e^{i \alpha I(M,C)}$\footnote{Our conventions here are that $I(M,C)$ is always an integer, as we take $M$ and $C$ to be chains with coefficients in $\zz$ (as opposed to $\rr$), so that the parameter $\alpha \in \rr/ 2\pi \zz$ must also be given to specify the holonomy of Wilson operators intersecting $M$.}.

In \eqref{states}, $|0\rangle$ is a reference state whose Wilson operators on the boundary $W_{C\subset \p X}$ all have $\langle W_C\rangle \in\rr$, which sets the reference for trivial holonomy. Such a reference choice can always be set for any given $A$ by re-defining the Wilson operators as $W_C \mapsto W_C\exp(i\int_C r(x))$ where $r(x)$ is some function which cancels the holonomy of $A$. This rescaling allows us to fix the values of $W_C$ in $|0\rangle$ to provide a reference state. Of course, only the relative phases of Wilson operator expectation values are physical (i.e. for a fixed $M$, the states $|\alpha,M\rangle$ form a $U(1)$ torsor), and as with regular $p=0$ SSB, asking which exact superselection sector one is in is not meaningful question. 

Finally, we point out that with these boundary conditions, we are unable to select ground states which are simultaneously in definite electric and magnetic superselection sectors (essentially, since $A$ and $\wt A$ can be thought of as Fourier transforms of one another). We construct ground states in a definite electric superselection sector by applying $e^{i\alpha Q(M)}$ to $|0\rangle$ and those in a definite magnetic superselection sector by applying $e^{i\beta \wt Q(N)}$ to $|\wt 0\rangle$ (here $|\wt 0\rangle$ is a reference state for the magnetic symmetry), but we cannot simultaneously specify both supersecltion sectors, since that would require boundary conditions that fix both $A|_{\p X}$ and $\star F|_{\p X}$, which is not allowed. Indeed, if we choose magnetic boundary conditions then acting with $Q(M)$ on an initial spatial slice $\Sigma$ gives
\be Q(M) = \frac{1}{g^2} \int_\Sigma (\star F)|_{\p X} \wedge \wh M =0,\ee 
and hence we cannot construct different electric symmetry-breaking states. Likewise, if we choose electric boundary conditions then 
\be \wt Q(N) = \frac{1}{2\pi}\int_\Sigma F|_{\p X}\wedge \wh N =0.\ee

From a more physical perspective, we can note that such boundary conditions would be those between vacuum and simultaneously an electric condensate (superconductor) and a monopole condensate, which is impossible since condensed electric charges confine magnetic ones and vice versa. 

One may think that the $p$-form symmetry transformation $A\mapsto A + \lambda$ is merely a shift in integration variables, and therefore any operators (viz. Wilson operators) which transform nontrivially under it must have zero expectation value \cite{deligne1999quantum}. Because of boundary conditions however, this is usually not the case. To illustrate this, consider a spacetime of the form $X=\Sigma \times \rr$, on which we impose electric boundary conditions by fixing $A|_{\p X}$. Perform a $p$-form symmetry transformation by shifting $A\mapsto A + \lambda$, where $\lambda$ has nontrivial holonomy along some submanifold $C$ of a certain time slice. Consider the effect of this shift on the operator $W_C$. As emphasized before, this shift is a {\it global} symmetry: since $d\lambda=0$, the holonomy of $\lambda$ around $C$ 
is independent of $t$, and so in particular $\lambda$ {\it cannot} go to zero at either end of the time interval. Thus the shift $A\mapsto A + \lambda$ must change the boundary conditions that we impose on $A|_{\p X}$, meaning that $A\mapsto A + \lambda$ is not a trivial shift in integration values, but rather a global symmetry. 

For example, in the archetypal case when $\Sigma = S^1\times \rr$ is a tube, we consider the transformation $A\mapsto A + (2\pi \alpha / L)d\theta$, where $L$ is the circumference of the $S^1$ and $\theta$ is the coordinate along the $S^1$. This is not pure gauge, and its holonomy around the $S^1$ is constant in time (and cannot be changed by a trivial gauge transformation). It changes the boundary conditions $A_\theta|_{\p X}$ at both the spatial boundary $\p\Sigma\times \rr$ and at $\Sigma \times \{\pm \infty\}$; therefore it is not a trivial shift of integration variables. 

One caveat: we will see in Section \ref{gauge_fixing} that one may write a generic $p$-form symmetry parameter as $\lambda=d^\da \beta + \omega$, where $\omega$ has nontrivial holonomy around some $C \subset Z_p(X,\p X)$, but is such that $\omega|_{\p X} =0$. In this case, $A\mapsto A+\omega$ is indeed just a shift in integration variables, and so if $\int_C \omega\neq0$, then $\langle W_C\rangle=0$. Such $\omega$ are measured by the relative cohomology group $H^p(X,\p X)$\footnote{Note that nontrivial elements in $H^p(X,\p X)$ can be globally exact, as long as they are the exterior derivative of a $(p-1)$-form that does not vanish on $\p X$.}. For $X=\Sigma \times \rr$ or $X=\Sigma \times I$, then we can apply the Kunneth formula for relative cohomology \cite{hatcher2002algebraic} and the long exact sequence of the pair $(I,\p I)$ to compute the relative cohomology as
\be \label{relative_coho_iso} H^p(X,\p X) \cong H^{p-1}(\Sigma,\p\Sigma).\ee

If we take $C\subset \Sigma \times \{t\}$ things simplify however, since no such $\omega$ exist that shift $W_C$ nontrivially. This is for the same reason as given above: if $\omega$ has nontrivial holonomy around $C$, then since it is closed, it must be non-zero at either end of the time interval, which is a contradiction since $\omega|_{\p X}=0$ by assumption\footnote{If we let $C$ not be embedded in a time slice, examples of such troublesome $\omega$ are easy to construct. For example, if $X = S^1\times I$, we can take $\omega = dg$, where $g(\theta,0)=a,g(\theta,1)=b$, and $a\neq b$. Then $\omega|_{\p X}=0$ and $d\omega=0$ but $\omega \not\in d\Omega^{p-1}_D(X)$, implying $\omega$ defines a nontrivial class in $H^1(X,\p X)$. If $C = (0,t)$ is a line running the length of the cylinder, $W_C$ shifts as $W_C\mapsto W_Ce^{i(b-a)}$ under the change of variables $A\mapsto A+\omega$, and so $\langle W_C\rangle =0$. However, if we take $C$ to be contained within a spatial slice, $W_C$ is invariant under the shift, and can have non-zero expectation value.}. 

\subsection{Ward identities} \label{ward_ids}

To elaborate on the role of boundary conditions, we can write down Ward identities for higher form symmetries. For a related discussion of the $p=1$ case, see \cite{avery2016noether}.

The expectation value of an operator $\mco$ is 
\be \langle \mco \rangle_{\Psi_{\p X}} = \int \frac{\mcd A}{\vol \mcg_0}\Psi_{\p X}[A]\mco  e^{iS},\ee
where $\Psi_{\p X}$ is the boundary wavefunctional\footnote{In the case where $\p X = \Sigma_i \cup \Sigma_f$, we would usually write this as $\Psi_{\p X} = \Psi_i[A]\Psi^*_f[A]$.} that controls the boundary conditions of the theory, and the subscript on the expectation value reminds us of the dependence of the expectation value on the boundary conditions. $\Psi_{\p X}$ is responsible both for setting boundary conditions on the initial and final time slices and for fixing the boundary conditions on the timelike component of $\p X$. 

In the path integral, we have written $\mcg_0$ for the group of gauge transformations that act trivially on the boundary conditions. This is naively $\mcg_0 = d\Omega^{p-1}_D(X)$, but as discussed in Section \ref{gauge_fixing} this is not quite correct if $p>1$, due to gauge-of-gauge transformations. The remaining nontrivial gauge transformations, i.e. those in $\mcg/\mcg_0$ where $\mcg$ is the group of all gauge transformations, are not integrated over in the path integral as they do not preserve the boundary conditions. 

We get the Ward identity by making a variation 
\be \label{deltaA} A\mapsto A + \epsilon\unit_R\lambda,\ee 
where $\lambda\in Z^p(X)$ is a flat $p$-form, $\epsilon$ is infinitesimal, and $\unit_R$ is a characteristic function that controls where on the 
spacetime manifold the symmetry action is being applied, defined so that $\unit_R(x) = 1$ if $x\in R$ and $\unit_R(x)=0$ otherwise, with $R$ some $D$-dimensional subregion of $X$. 
The Ward identity is then 
\be \label{ward_id} \langle \delta \mco \rangle_{\Psi_{\p X}} + \langle \mco \rangle_{\delta\Psi_{\p X}} + i\langle \delta S \mco \rangle_{\Psi_{\p X}}=0.\ee  
Since the $p$-form symmetry parameter $\lambda$ is closed, we have $d(\unit_R\lambda) = d\unit_{R}\wedge \lambda = \wh{\p R}\wedge\lambda$. Thus, the $\delta S$ contribution is
\be \delta S = -\frac{\epsilon}{g^2} \int_{\p R} \star F \wedge \lambda,\ee
which is proportional to the charge operator $Q(\wh \lambda)$. 

The second term in \eqref{ward_id} will make a contribution if the region $R$ has support on the boundary of the spacetime manifold, ${\rm Supp}(R)\cap \p X \neq 0$. Note that if we take $R=X$ to be the entire spacetime manifold then the $\delta S$ term is zero since when $R=X$ we are performing a global symmetry transformation. In this case, the expectation value of $\delta\mco$ is completely determined by the shift in boundary conditions. This is why these boundary terms are crucial: if we did not include them, spontaneous symmetry would not be possible. Relatedly, we see that if we work on a compact spacetime with $\p X = 0$, the Ward identity with $R=X$ tells us that $\langle \mco \rangle = 0$; without external fields SSB can only take place on non-compact manifolds (of course, even if $\p X\neq 0$, if the boundary wavefunctional $\Psi_{\p X}$ is symmetric so that $\delta \Psi_{\p X}=0$ then $\langle \mco \rangle_{\Psi_{\p X}} = 0$, and SSB is impossible). 

As mentioned in the previous subsection, our procedure for constructing different ground states is to apply the translation operator $e^{i\epsilon Q(M_0)}$ to a given reference state $|0\rangle$, where $M_0$ is the Poincare dual of the boundary field configuration $A_0\equiv A|_{\p X}$. Thus for small variations away from the reference state $|0\rangle$, the first order variation in the boundary wavefunctional under the shift \eqref{deltaA}, which sends $A_0 \mapsto A_0 + \lambda$, is 
\be \delta |\Psi_{\p X}\rangle = \delta\, \exp \left( i \frac{\epsilon}{g^2} \int_{\p X} \star F \wedge A_0\right) |0\rangle = i \frac{\epsilon}{g^2} \int_{\p X} \star F \wedge \lambda\,  | \Psi_{\p X}\rangle.\ee

Setting $R=X$, we then get the global Ward identity 
\be i\epsilon \left\langle\mco \frac{1}{g^2} \int_{\p X} \star F \wedge \lambda\right\rangle_{\Psi_{\p X}} = \langle \delta\mco\rangle_{\Psi_{\p X}}.\ee
In particular, if $\mco = W_C$ is a Wilson operator, this reads
\be \label{wilson_ward} \left\langle W_C \frac{1}{g^2} \int_{\p X} \star F \wedge \lambda\right\rangle_{\Psi_{\p X}}  =  \left(\int_C \lambda\right) \langle W_C\rangle_{\Psi_{\p X}}.\ee
This is essentially the infinitesimal version of the statement that $e^{i \alpha Q(M)}$ shifts $W_C$ by $e^{i\alpha I( M,C)}$, where $\lambda = \wh M$.
Note that the RHS of \eqref{wilson_ward} is only non-zero if $\int_C\lambda \neq 0$, i.e. if the charge $I(\wh \lambda,C) \in \zz$ of the Wilson operator $W_C$ is non-zero. 

It can be useful to draw pictures to visualize the statement of the Ward identity. An easy case to draw is regular $p=1$ gauge theory in $D=3$ Euclidean space. Consider the symmetry $A\mapsto A + \lambda$ with $\lambda = \rho(x) dx$, where $\rho(x)$ is a bump function centered on $x=0$. The Poincare dual of $\lambda$ in the full spacetime $X$ is the $yt$ plane, while the Poincare dual $\wh \lambda$ taken within a spatial slice is the $y$ axis, and so the charge operator $Q(\wh \lambda)$ measures the electric flux passing through the $y$-axis. 

The Ward identity for this symmetry can be schematically written as 
\be \includegraphics{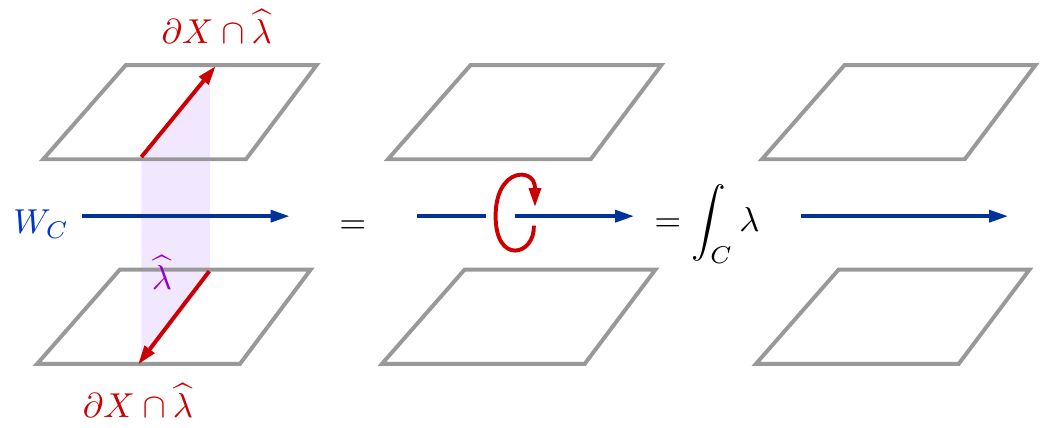}. \ee
The blue line marks the support of $W_C$, while the red lines indicate the support of $(\star F \wedge \lambda)|_{\p X}$. The gray parallelograms are meant to depict sections of initial and final spatial slices. 
The pinkish rectangle filling the $yt$ plane in the first figure is the Poincare dual of $\lambda$ with respect to the full spacetime $X$, and its restriction to the boundary of $X$ is indicated by the red lines. 
In the second step, we have used current conservation to deform the support of the charge operator (namely $\p X \cap \wh \lambda$) into a small loop linking $C$: this deformation is allowed
as long as we don't cross any operators charged under the symmetry (viz. the Wilson line). In the last step, we remove the loop at the expense of a factor $\int_C\lambda = I(C,\wh \lambda)$, which is the charge of $W_C$ and comes from passing the red loop ``through'' the Wilson line.

\subsection{Gauge fixing \label{gauge_fixing}}

We now turn to the issue of gauge fixing. Together with our choice of boundary conditions, the gauge-fixing process should eliminate all local degrees of freedom and leave us with a boundary value problem with a unique solution on the classical equations of motion, up to possible topological obstructions. 

We will choose to work with the class of gauge-fixing functions 
\be \label{gauge_cond} d^\da A - f=0\ee 
for some co-closed $(p-1)$-form $f$, 
which is convenient from a technical perspective since it allows us to take advantage of results in Hodge theory. 
Any nontrivial higher $p$-form symmetries should leave this gauge-fixing condition invariant, since they are not trivial gauge transformations. 

First, some preliminaries. On a Riemannian manifold $X$ with non-empty boundary, we have a Hodge decomposition of the $p$-forms on $X$, namely \cite{cappell2006cohomology}
\be \label{hodge_decomp} \Omega^p(X) = d\Omega^{p-1}_D(X) \oplus d^\da \Omega^{p+1}_N(X) \oplus (\ker d \cap \ker d^\da \cap \Omega^p(X)),\ee
where the direct summands are orthogonal under the inner product $\langle \alpha,\beta\rangle = \int_X \alpha \wedge \star \beta$. Since $d$ commutes with the inclusion $i_\p : \p X \ra X$, if $\alpha$ is Dirichlet then so is $d\alpha$, implying $d\Omega_D^{p-1}(X) \subset \Omega^p_D(X)$. 
Likewise, because the Hodge dual $\star$ exchanges Dirichlet and Neumann boundary conditions, if $\alpha$ is Neumann then so is $d^\da \alpha$, and so $d^\da \Omega^{p+1}_N \subset \Omega_N^p(X)$. 

Trivial gauge transformations, which are the ones we need to eliminate when performing the path integral, must satisfy two conditions. First, they must be given by an exact $p$-form $d\gamma$. 
Secondly, since they must preserve the boundary conditions (which we will assume to be given by fixing $A|_{\p X}$), we require $(d\gamma)|_{\p X}=0$. Note that if $(d\gamma)|_{\p X}\neq0$ we do not regard $d\gamma$ as a trivial gauge transformation, even though such a transformation changes $A|_{\p X}$ by an exact form. This is because in order to have a well-defined boundary value problem and in order for the symplectic form \eqref{symplectic_form} to be invertible, $A|_{\p X}$ must be exactly fixed by the boundary conditions, not just fixed up to an exact form. We cannot fix these exact forms on the boundary away with a gauge-fixing procedure, since solving a gauge condition like $d^\da A=0$ requires knowledge of $A|_{\p X}$ in the first place. Therefore, $A|_{\p X}$ needs to be precisely fixed, and so all trivial gauge transformations must have $(d\gamma)|_{\p X}=0$. 

Actually, as is usually done in the literature, we will make the slightly stronger requirement that $\gamma|_{\p X}=0$. This means that $\gamma$ with $(d\gamma)|_{\p X}=0$ but $\gamma|_{\p X}\neq0$ will not be regarded as trivial (local) gauge transformations --- this is because such a $\gamma$ differs from one that does vanish on the boundary by a global symmetry, which we do not gauge-fix away. The simplest example is when $\gamma$ is a 0-form, with $\gamma|_{\p X}=c$ a constant. Then $\gamma' = \gamma-c$ differs from $\gamma$ by a global symmetry and $\gamma'|_{\p X}=0$, so that $d\gamma'$ is a trivial gauge transformation, but not $d\gamma$.

Recapitulating, the group of trivial gauge transformations on $A$ is given by the group $d\Omega_D^{p-1}(X).$ 
Note that if we had instead fixed magnetic boundary conditions by imposing $\star F|_{\p X}=0$, the group of trivial gauge transformations on $A$ would be the full $d\Omega^{p-1}(X)$ (without the Dirichlet constraint), since with magnetic boundary conditions $A|_{\p X}$ is unfixed.  

We need to check that $d^\da A - f = 0$ is a good gauge-fixing condition, i.e. we need to check that if we are given a field configuration with $d^\da A -f \neq 0$, then we can always find a trivial gauge transformation $d\gamma \in d\Omega_D^{p-1}(X)$ such that $d^\da(A - d\gamma)-f=0$. 
In order for the gauge condition to be viable, there must be some $A_0$ that solves $d^\da A_0 - f =0$, and so by Hodge-decomposing $A_0$, we see that $f=d^\da d \alpha_0$ for some $\alpha_0 \in \Omega^{p-1}_D(X)$. 
Now consider an arbitrary $A$ with 
\be d^\da A - f = g.\ee
Then after Hodge-decomposing $A = d\alpha + d^\da \beta + \omega$ with $\alpha \in \Omega^{p-1}_D(X)$, we see that $g=d^\da d (\alpha - \alpha_0)$. 
Thus we define $A' = A + d\gamma$ with 
\be \gamma = \alpha - \alpha_0 + d\varepsilon,\ee 
with $d\varepsilon\in d\Omega^{p-2}_D(X)$ representing the gauge redundancy in $\gamma$ itself. Since $d\gamma \in \Omega^p_D(X)$, $A$ and $A'$ are gauge equivalent, and since $d^\da A'-f=0$, \eqref{gauge_cond} is indeed a good gauge-fixing condition.  

Finally, we need to check that $d^\da A-f=0$ fixes the gauge completely. That is, we need to check that there are no nontrivial solutions to $d^\da d\gamma=0$, with $d\gamma \in d\Omega^{p-1}_D(X)$ a trivial gauge transformation. Indeed this is the case, which we can see by decomposing the closed $p$-forms on $X$ as
\be \label{closed_hodge} Z^p(X) = (\ker d \cap \ker d^\da \cap \Omega^p(X) ) \oplus d\Omega^{p-1}_D(X).\ee
Now a trivial gauge transformation $d\gamma$ is certainly in $Z^p(X)$, but by assumption it is also in $d\Omega^{p-1}_D(X)$, and so from the above we see that it is orthogonal to $\ker d^\da$. Hence if $d^\da d\gamma=0$ then $d\gamma=0$, meaning that $d^\da A-f=0$ fixes the gauge completely. 

We now turn to the uniqueness of the solution of the classical equations of motion. Given $A|_{\p X}$, suppose we have two field configurations $A_1,A_2$, each satisfying the gauge-fixing condition and the classical equation of motion $d^\da d A=0$, and with boundary conditions $A_1|_{\p X} = A_2|_{\p X} = A|_{\p X}$. Then $\bar A = A_1 - A_2$ satisfies $d^\da \bar A = d^\da d \bar A =0,$ with $\bar A|_{\p X}=0$. These conditions mean that $\bar A$ is a Harmonic $p$-form that vanishes on $\p X$ (and also imply $d\bar A=0$). Such forms are isomorphic to the relative de Rham cohomology, and so the obstruction to uniquely solving the classical equations of motion is
\be {\rm Harm}_D^p(X) \cong H^p(X,\p X),\ee
which is generically non-zero and represents a topological obstruction to solving the equations of motion. 
For $X=\Sigma \times I$ or $\Sigma \times \rr$, this obstruction can be computed with the help of \eqref{relative_coho_iso}. In particular, since $H^0(\Sigma,\p \Sigma)$ is always trivial, ${\rm Harm}_D^1(X)=0$. 

Note that if we also required a boundary condition on $\star A|_{\p X}$, then we must also have $\bar A \in \Omega^p_N(X)$. The set of closed and co-closed forms with Neumann boundary conditions is isomorphic to $H^p(X)$, and so $\bar A \in H^p(X)$. But in fact 
\be H^p(X) \cap H^p(X,\p X) = 0,\ee
and so in such a case we have $\bar A = 0$, implying the classical problem has a unique solution. 

In any case, if $H^p(X,\p X) \neq 0$ and we place no boundary conditions on $\star A|_{\p X}$, then we can divide up $X$ into patches $X_i$ with $H^p(X_i,\p X_i)=0$, solve the classical problem on each patch, and then glue the patches together by matching gauge fixing conditions on the overlaps. 

Since $p$-form symmetries must respect the gauge-fixing condition, we can identify higher-form symmetries by searching for $p$-forms $\lambda$ with $d\lambda = 0$ and $d^\da \lambda = 0$, modulo trivial gauge transformations. The requirement $d^\da \lambda=0$ was imposed because of our choice of gauge, but in order for these symmetries to be true symmetries their classification must be independent of the choice of gauge fixing. However, it turns out that all higher-form symmetry parameters can be chosen to satisfy $d^\da \lambda=0$, regardless of our choice of gauge fixing, essentially for the same reason that $d^\da A=0$ is a good gauge condition. Indeed, consider a candidate $p$-form symmetry transformation parameter $\lambda \in Z^p(X)$. From the Hodge decomposition of closed $p$-forms \eqref{closed_hodge} we see that we may write $\lambda = d\gamma + \lambda'$, where $d\gamma \in d\Omega^{p-1}_D(X)$ is a trivial gauge transformation and $d^\da\lambda'=0$. 
Therefore, since we are only interested in gauge equivalence classes of $p$-form symmetry parameters, we may set $d^\da\lambda=0$ regardless of our choice of gauge-fixing term. 

Thus, to look for $p$-form symmetries we need to examine the $p$-forms that are contained within $\ker d \cap \ker d^\da$. This space decomposes as 
\be \ker d \cap \ker d^\da \cap \Omega^p(X) = (\ker d \cap d^\da \Omega^{p+1}(X)) \oplus (\ker d \cap \ker d^\da \cap \Omega^p_D(X)).\ee
The latter part of the direct sum contains only forms that do not change the boundary data, and so any $p$-form symmetry parameters that {\it do} act on the boundary data are all closed and co-exact. We can also write the latter summand as 
\be \ker d \cap \ker d^\da \cap \Omega_D^p(X) = H^p(X,\p X),\ee
meaning that we can write a generic $p$-form symmetry parameter as 
\be \label{psym_decomp} \lambda = d^\da \beta + \omega,\ee 
where $dd^\da \beta = 0$ and $(d^\da \beta) |_{\p X}\neq 0$, and where $\omega \in H^p(X,\p X)$ preserves the boundary conditions. If we require that $\omega$ preserve $\star A|_{\p X}$ as well, then $\star \omega|_{\p X}=0$ and hence $\omega \in H^p(X)$. Because $H^p(X,\p X) \cap H^p(X) = 0$, in this case we may set $\omega = 0$. 

In any case, the contribution from $\omega$ is actually irrelevant for examining the issue of symmetry breaking, for the reason discussed in Section \ref{ward_ids}. Indeed, if $W_C$ transforms nontrivially under the shift $A\mapsto A+\omega$ then we must have $\langle W_C\rangle =0$, because $\omega \in \Omega^p_D(X)$ means that shifting $A$ by $\omega$ is a trivial shift in integration variables, and hence any operators that transform under it must have zero expectation value. 

Finally, we briefly comment on how the gauge-fixing works in the path integral. 
We first split up the gauge field as 
\be A=A_{cl} + A_q,\ee 
where $A_{cl}$ is a solution to the classical equations of motion $d^\da d A_{cl}=0$. The fields have the following boundary conditions: 
\be A_{cl}|_{\p X} = A|_{\p X},\qquad A_q|_{\p X} = 0.\ee
Since the boundary conditions are imposed on the classical part, the classical part dictates how the symmetry is spontaneously broken. 

The off-diagonal terms in the action vanish, since $A_q|_{\p X}=0$ allows us to write 
\be \int_X dA_q\wedge \star dA_{cl} = \int_X A_q \wedge \star d^\da d A_{cl} = 0.\ee
We can then use $\int_X dA_q \wedge \star dA_q = \int_X A_q \wedge \star d^\da dA_q$ to write 
\be S = S[A_{cl}] + \frac{1}{2g^2} \int_X A_q \wedge \star d^\da dA_q.\ee
$d^\da d$ isn't invertible acting on $p>0$ forms, and so we need to add a gauge-fixing term.
Choosing the gauge-fixing function to be $d^\da A -f=0$ and averaging over all co-exact $f$ produces\footnote{Defining a measure for a path integral over only co-exact $f$ is slightly subtle, but the subtitles only manifest themselves in the counting of the number of ghosts needed to impliment the gauge-fixing, and as such will not play an important role in what follows.} 
\be S = S[A_{cl}] + \frac{1}{2g^2}\int_X A_q\wedge\star\left(d^\da d + \frac{1}{\alpha}dd^\da\right)A_q.\ee
Choosing $\alpha=1$ gives the analogue of the Feynman gauge, with the $A_q$ term becoming $A_q\wedge\star\Delta A_q$. As usual, the gauge-fixing term also requires the addition of a pair of fermionic $(p-1)$-form ghosts $\xi$, with action $\xi \wedge \star d^\da d \xi$ and Dirichlet boundary conditions $\xi|_{\p X}=0$. 

We have only mentioned the gauge fixing of $A$, but when $p>1$ we also have to deal with gauge-of-gauge transformations \cite{siegel1980hidden}, since the gauge transformation $\gamma$ in $A \mapsto A + d\gamma$ is itself only defined up to an element of $d\Omega^{p-2}_D(X)$. Thus, when we mod out by all gauge parameters $\gamma \in \Omega^{p-1}_D(X)$ we are actually killing too many degrees of freedom, since not all of the gauge parameters are independent. We also need to gauge fix the $(p-1)$-form fields $\gamma$, which we do by working with the same form of gauge condition that we imposed on $A$, namely $d^\da \gamma - f'=0$. 

$\gamma$ appears in the action in which $A$ has been gauge-fixed in the form $\int_X \gamma \wedge \star (d^\da d)^2 \gamma.$
If $p>1$ then $d^\da d$ is not invertible acting on $(p-1)$-forms, and so we are prompted to add a gauge-fixing term of the form $\int \gamma \wedge \star (dd^\da)^2\gamma$ so that the full term with $\gamma$ becomes
\be \int_X \gamma \wedge \star[(dd^\da)^2 + (d^\da d)^2]\gamma = \int_X \gamma \wedge \star \Delta^2\gamma.\ee 
When acting on $\gamma$, the Laplacian $\Delta$ is invertible modulo the topological obstruction $H^{p-1}(X,\p X)$. Adding this gauge-fixing term necessitates adding several additional ghosts for $\gamma$, in addition to the ghosts required for the gauge-fixing of $A$\footnote{These ghosts are bosonic since they {\it restore} degrees of freedom that were killed by eliminating the longitudinal mode of $A$, which itself has a gauge redundancy.}.

These gauge-of-gauge transformations are themselves still not independent if $p>2$: they are only defined up to an element of $d\Omega^{p-3}_D(X)$. Thus we must add a tower of gauge-fixing terms and ghosts, continuing this procedure until we reach the bottom of the tower where the gauge parameters are zero-forms \cite{siegel1980hidden}. The fields at each level of the tower are all fixed in the same way that the $p$-form component of the gauge symmetry is fixed, with the questions about boundary conditions, the goodness of the gauge fixing function, and the uniqueness of its solutions having the same answers as in the case for $A$ outlined above. We will omit further details, since questions regarding SSB are addressed in the same way for each level of the tower.

\section{Higher symmetries on cohomologically trivial manifolds and asymptotic symmetries \label{asymp_syms}}

In this section, we briefly discuss higher-form symmetries on manifolds with trivial cohomology and possible relationships between higher-form symmetries and asymptotic symmetries in gauge theories, which have recently been the subject of much interest \cite{seraj2017multipole,strominger2014asymptotic,avery2016noether,he2014new,strominger2017lectures}. Most of the discussion on asymptotic symmetries in the literature has focused on particular choices of spacetimes and particular ways of dealing with infinity, with each choice being handled in a slightly different manner (see e.g. \cite{seraj2017multipole,he2014new,strominger2017lectures,campiglia2017asymptotic}). Our hope is that while the existence of nontrivial asymptotic symmetries may generically depend on these choices, the general idea of higher symmetries may provide a framework for thinking about asymptotic symmetries (at least in Abelian gauge theories) in a slightly more cohesive way. Some similar remarks for the $p=1$ case were noted in \cite{rasmussen2017gapless}; we also note the recent \cite{afshar2018asymptotic}, in which types of asymptotic symmetries for $p$-form theories in Minkowski space are discussed.

\begin{figure}
\centering
\includegraphics{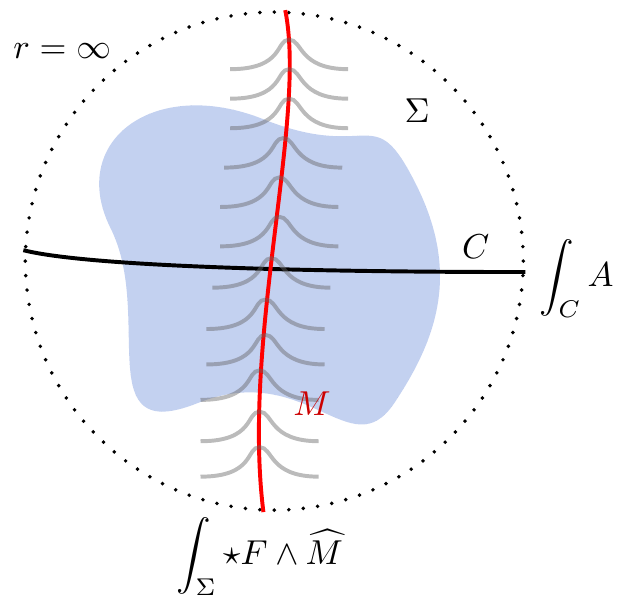}
\caption{\label{infinite_shift_fig} A two-dimensional spatial slice $\Sigma$, with the dotted circle representing $r=\infty$. The Wilson line $W_C=\exp(i\int_C A)$ (in black) is charged under a 1-form symmetry generated by $\star F$ integrated along the red curve $M$. This shifts the gauge field by $\wh M$, which is the bump function Poincare dual to $M$ and is shown schematically in grey.}
\end{figure}

For certain choices of boundary conditions (or asymptotic fall-off conditions), Abelian gauge theories on unbounded manifolds possesses an infinite-dimensional group of asymptotic symmetries, which are ``large'' gauge transformations that have support at infinity (see e.g. \cite{seraj2017multipole,strominger2014asymptotic,avery2016noether,he2014new,strominger2017lectures}). These asymptotic symmetries are a subset of the group of higher-form symmetries of the theory, namely the ones which shift the gauge field by exact $p$-forms which have support at infinity, and which exist even when $H^p(X)=0$, which we will assume for the remainder of this section. 
The objects charged under asymptotic symmetries are Wilson operators $W_C$, with $\p C$ supported at infinity. An example of the general idea for the case of $D=3,p=1$ is shown in Figure \ref{infinite_shift_fig}, where we have drawn a Wilson operator $W_C$ and a charge operator on a spatial slice $\Sigma = \rr^2$. 

To determine if there are any asymptotic symmetries, we need to figure out if there are any gauge transformations $A \mapsto A+ d\varepsilon$ which act nontrivially on the physical data of the theory (viz. the boundary conditions). As in previous sections, the boundary conditions will be imposed on the gauge fields at $\p X$ (i.e. at infinity). By boundary conditions at $\p X$, we really mean asymptotic fall-off conditions on $A$: these typically require the radial component of $A_r$ to vanish with a certain power of $1/r$, while allowing the tangential components at infinity $A_{||}$ to be an arbitrary closed form. It is often convenient to choose Cauchy surfaces $\Sigma$ orthogonal to the radial direction, in which case $A_{||}$ is specified as part of our boundary conditions, and so gauge transformations which shift $A_{||}$ are higher symmetries. 

Suppose $A\mapsto A + d\varepsilon$ is a nontrivial asymptotic / higher symmetry transformation. $\varepsilon$ cannot be compactly supported\footnote{It is thus tempting to regard higher symmetries on non-compact manifolds as being parametrized by $L^2$ cohomology, but imposing the restriction that $A$ be $L^2$ (as opposed to just $F$) is an unphysically strong restriction on the gauge field.}, 
and so a necessary condition for $d\varepsilon$ to generate a global symmetry is that it be exact in $\Omega^p(X)$ but not in $\Omega^p(X,\p X)$. The Poincare dual statement to this is that the charge $Q(\wh{d\varepsilon})$ can only generate a nontrivial asymptotic symmetry if the $p$-manifold $\wh{d\varepsilon}$ has non-empty boundary with $\p (\wh{d\varepsilon})\subset\p X$, so that $\wh{d\varepsilon}$ is in $Z_{D-p}(X,\p X)$ but not in $Z_{D-p}(X,\rr)$. 

If $d\varepsilon$ is not a trivial local gauge transformation, it should respect the gauge condition $d^\da d\varepsilon =0$. As we did when discussing gauge-fixing, we can decompose exact $p$-forms as 
\be d\Omega^p(X) = d\Omega^p_D(X) \oplus (d\Omega^p(X) \cap \ker d^\da),\ee 
and so if $d\varepsilon|_{\p X}\neq0$, the gauge condition $d^\da d\varepsilon=0$ is automatically satisfied. 

We should point out that often when studying asymptotic symmetries, one only focuses on $\varepsilon|_{\p X}$. However, here we stress that we are always thinking of $\varepsilon$ as a $(p-1)$-form defined throughout the full spacetime. Of course, $\varepsilon$ in the interior of $X$ can be obtained from $\varepsilon|_{\p X}$ by integrating it against the propagator on $(p-1)$-forms, but this is only possible if we have chosen a good gauge condition. We stress that we are imposing this gauge condition globally throughout $X$, not just asymptotically. 

In most of the literature, the Cauchy surface $\Sigma$ is taken to be past or future null infinity. For regular $p=1$ Maxwell theory in $D=4$ Minkowski space, one often fixes the boundary conditions so that the radial component $A_r$ falls off to zero at infinity, while the angular components $A_z,A_{\bar z}$ (the ones tangent to the $S^2$ at infinity) are constrained only by the requirement that the field strength on the boundary vanish (see e.g. \cite{strominger2017lectures}). 

Now consider the Wilson line $W_C$, where the curve $C\subset \Sigma$ starts and ends at the points $z_i,z_f$. Under the 1-form asymptotic symmetry $A\mapsto A + d\varepsilon$ where $\varepsilon$ is time-independent and $d\varepsilon|_{\Sigma}\neq0$, $W_C$ transforms as 
\be W_C \mapsto W_C e^{i[\varepsilon(z_f,\bar z_f) - \varepsilon(z_i,\bar z_i)]}.\ee
The charge operators for this symmetry are 
\be Q(\wh{d\varepsilon}) = \frac{1}{g^2}\int_{\Sigma} \star F\wedge d\varepsilon = \frac{1}{g^2}\int_{\p \Sigma} \star F \wedge \varepsilon,\ee 
where $\varepsilon$ can be an arbitrary function. This expression for the charge operator is precisely the same as the one for asymptotic symmetries appearing in e.g. \cite{strominger2017lectures, he2014new,seraj2017multipole,avery2016noether,strominger2014asymptotic}. 

As an example, let $\mcs^2$ denote the sphere at infinity, and take $\varepsilon|_{\Sigma}$ to be the time-independent 0-form that vanishes on the southern hemisphere of $\mcs^2$ and equals 1 on the northern hemisphere. The Poincare dual $\wh \varepsilon$ is then the northern hemisphere of $\mcs^2$, where the Poincare dual is taken in $\Sigma$. Thus $d\varepsilon$ is a bump function supported on the equator of $\mcs^2$, and its dual $\wh{d\varepsilon}$ is the 2-manifold $\mcs^1\times \rr$, where $\mcs^1$ is the equator of $\mcs^2$.
We can then determine $\varepsilon$ on all of $X$ from its value on $\Sigma$ by using the gauge-fixing condition $d^\da d \epsilon =0$, which is imposed throughout the interior $X$.

With this choice of $\varepsilon$, any Wilson line $W_C$ which connects the two hemispheres of $\mcs^2$ is charged under this asymptotic symmetry. Specifically, acting with with $e^{i\alpha Q(\wh{d\varepsilon})}$ shifts $W_C$ by $e^{ i\alpha}$ if $C$ runs from the southern hemisphere to the northern hemisphere (or by $e^{-i\alpha}$ if $C$ runs the other way), while it leaves $W_C$ invariant otherwise. 

More generally, for $p$-form theories on any unbounded manifold, we have charged operators $W_C=\exp(i\int_C A)$, where $C$ is any $p$-dimensional submanifold whose boundary lies at infinity. 
There are an infinite number of $p$-form symmetries under which the collection of $W_C$ are charged, one for each equivalence class of relatively closed $(D-p-1)$-manifolds $M$ with $\p M$ supported at infinity, where the equivalence relation is given by homotopies which preserve $\p M$. Using Poincare duality to write $M=\wh{d\varepsilon}$, the charge operators for these symmetries are $Q(\wh{d\varepsilon})=\frac{1}{g^2}\int_\Sigma \star F\wedge \varepsilon$.

\section{The higher Coleman-Mermin-Wagner theorem} \label{CMW}

The higher Coleman-Mermin-Wagner theorem is as follows \cite{gaiotto2015generalized}:
\[\ba &\text{\it continuous $p$-form symmetries in $D$ spacetime dimensions are never broken if $p\geq D-2$.}\ea\]
In particular, taking $p=0$ gives the regular CMW theorem. When we set $p=1$, we have the statement that 1-form symmetries are never broken in three dimensions: as we will see, this is equivalent to the statement that pure $U(1)$ gauge theory is confining for $D=3$. 

Note that in the case of discrete symmetry the condition is modified to $p\geq D-1$. This allows us to reproduce the statement that topological order (phases were a discrete 1-form symmetry is spontaneously broken) can exist in three spacetime dimensions, but not in two \cite{chen2011classification}. 

We also see that while the higher CMW theorem allows us to interpret the photon as a Goldstone boson of a spontaneously broken 1-form symmetry for $D=4$, such an interpretation is not possible for $D=3$. This is similar to the statement that the compact boson in two dimensions, while a massless scalar, is not the Goldstone boson of any spontaneously broken symmetry. 

We will give a ``proof'' by showing that the charged operators for $p<D-2$ theories are allowed to acquire nonzero expectation values, while for $p\geq D-2$ they cannot. An easy case is for $p=D-1$, when $\star F$ is a $0$-form. The charge operators $Q(M)$ are then local, as $M$ is a single point. Since the charge operators (which connect different putative symmetry-breaking vacua) are local, transitions between different vacua cost finite action can be accomplished locally, implying that there are no superselection rules and that spontaneous symmetry breaking is not possible. Thus, $p=D-2$ is at least an upper bound on the choice of $p$ for which SSB is allowed. 

\subsection{In the continuum (no monopole operators) \label{continuum} } 

Now first turn to an analysis of SSB and the CMW theorem in the continuum, with no magnetic matter, so that $dF=0$ and the $(D-p-2)$-form magnetic symmetry is preserved.
Our goal is to compute expectation values of Wilson membrane operators in $p$-form theories and show how their behavior depends on the spacetime dimension (we are aware that similar computations have been performed by S. Grozdanov \cite{grozdanov2018talk}). 

The basic idea for evaluating $\langle W_C\rangle$ is to split up momentum integrals into modes lying along the $p$-submanifold $C$ and modes transverse to it. The momenta which generate longitudinal diffeomorphisms of the Wilson brane are unphysical, and so only the $D-p$ transverse modes are relevant. Thus the relevant momentum integral for computing $\langle W_C\rangle$ will be something like $\int d^{D-p}k \frac{1}{k^2}$. This integral is divergent when $D-p\leq 2$, which hints at the correctness of the higher CMW theorem\footnote{Thanks to Ryan Thorngren for emphasizing this argument to me.}.  

In more detail, in order to diagnose SSB we need to compute the expectation values 
\be \langle W_C\rangle =\left\langle\exp\left(i\int_C A\right)\right\rangle = \left\langle \exp \left(i \int_XA\wedge \wh C\right)\right\rangle,\ee
where as usual $\wh C$ is the Poincare dual of $C$ with respect to $X$\footnote{When $C$ is a boundary, we can also consider Wilson operators with different charges, namely $\exp(i\lambda\int_CA)$, where $\lambda$ is a real number. We cannot do this when $C$ is not a boundary however, since the resulting Wilson operator is not invariant under large gauge transformations which have nontrivial winding around $C$. When we have dynamical matter around Wilson operators with $\lambda \in \zz$ can be screened, but for the pure gauge theories we are studying this is not an issue.}.
The expectation value is (now in Euclidean signature)
\be \langle W_C \rangle = \int_X \mcd A \mcd \mu_{gh}\exp\left[ -\left( \int_X \frac{1}{2g^2}F \wedge \star F - i A \wedge \wh C\right) - S_{gf} - S_{gh}\right],\ee
where $S_{gf} \sim \frac{1}{\alpha}\int_X A \wedge \star dd^\da A$ is the gauge-fixing term.
$\mcd\mu_{gh}$ is the measure for all the ghosts and the ghost action $S_{gh}$ includes the ghosts and all of their gauge-fixing terms, which are needed if $p>1$ (see Section \ref{gauge_fixing}). The ghosts won't play a further role in the present discussion, and so we will omit them in what follows. 

After splitting the gauge field into $A=A_{cl}+A_q$ as in Section \ref{gauge_fixing}, we have
\be \langle W_C \rangle = \int {\mcd A} \, e^{-S[A_{cl}]} \, \exp\left[ -\int_X A_q \wedge \left(\frac{1}{2g^2}\star \left( d^\da d + \frac{1}{\alpha}d d^\da \right)A_q - i \wh C\right)\right].\ee
The $e^{-S[A_{cl}]}$ contribution is a constant that won't affect our analysis of the behavior of $|\langle W_C\rangle|$, since it is $C$-independent (and therefore we will drop it). 

We then eliminate the $A_q\wedge\wh C$ coupling by shifting the $A_q$ term by 
\be A_q\mapsto A_q - ig^2 \frac{1}{d^\da d + \frac{1}{\alpha}dd^\da}\star \wh C.\ee 
Such a shift is allowed, since $(d^\da d + \frac{1}{\alpha}dd^\da)^{-1}$ is invertible on $A_q$, owing to $A_q$'s boundary conditions. Furthermore, if $C$ meets $\p X$ transversely (or not at all), which we will assume, then such a shift leaves the boundary conditions $A_q|_{\p X}=0$ invariant (since in that case, $\wh C$ is Neumann, implying $\star \wh C$ is Dirichlet). 

This leaves us with a Gaussian path integral over $A_q$, producing a constant factor of $(\det \Delta)^{-1/2}$ which, being $C$-independent, we will omit. We then have
\be \langle W_C \rangle = \exp\left(-\frac{1}{2}g^2\int\wh C \wedge \star D  \wh C\right),\ee
where $D$ is the real-space propagator, i.e. $D= (d^\da d + \frac{1}{\alpha}dd^\da)^{-1}$. Writing this out in full, we have 
\be \ba\langle W_C \rangle & = \exp\Bigg(-\frac{1}{2}g^2\int_C \int_C dx^{\mu_1} \wedge \dots \wedge dx^{\mu_p} \wedge dy^{\nu_1}\wedge \dots\wedge dy^{\nu_p}\ D_{\mu_1\dots\mu_p\nu_1\dots\nu_p}(x-y)\Bigg),\ea\ee
where the integrals both run over the surface $C$. This expression is just the generalization of the self-energy of a current-carrying wire to higher dimensions. 

In order to run through examples, we will need expressions for $D$, the $p$-form propagator, in different dimensions. The result for the higher analogue of Feynman gauge will turn out to be very similar to the $p=1$ case (uninterested readers may skip ahead). 
Since the Hodge Laplacian $\Delta$ is invertible (putting aside topological restrictions coming from relative cohomology groups; if these do not vanish we must work on patches), we can construct the projector $P = d^\da d/\Delta$. This is a projector since 
\be P^2 = \frac{d^\da d d^\da d}{\Delta^2} = \frac{d^\da(\Delta-d^\da d)d}{\Delta^2} = \frac{d^\da d}{\Delta} = P,\ee
as $d^2=0$ and $\Delta d = d\Delta$. We then construct the propagator by assuming that $D$ is formed as a linear combination of the identity and $P$; we find 
\be D = \frac{1-\alpha}{\Delta} \frac{d^\da d}{\Delta} + \frac{\alpha}{\Delta}.\ee
Taking $\alpha=1$ gives the analogue of Feynman gauge. In flat space, the momentum-space propagator in Feynman gauge is then given by the very simple expression 
\be \label{general_propagator} D_{\mu_1\dots\mu_p}^{\nu_1\dots\nu_p}(k) = \frac{\delta_{\mu_1\cdots\mu_p}^{\nu_1\dots\nu_p}}{k^2},\ee
where as usual the delta function $\delta_{\mu_1\cdots\mu_p}^{\nu_1\dots\nu_p}$ is $1$ ($-1$) if the $\{ \mu_i\}$ are an even (odd) permutation of the $\{\nu_i\}$, and $0$ otherwise. 

There are two illustrative classes of Wilson operators $W_C$ that we will do some calculations for. The first is the case where $C$ is some copy of $\rr^p$ embedded within $\rr^D$. As mentioned in Section \ref{asymp_syms} these Wilson operators are gauge invariant since gauge transformations are not allowed to change the gauge field at infinity. 

The calculation is straightforward: we take $C$ to be a copy of $\rr^p$ inside $\rr^D$, which we think of as a $D$-dimensional box with side length $L \ra \infty$. We have 
\be \ba \label{big_sheet_int} \langle W_C \rangle & = \exp\left(-\frac{1}{2}g^2 \int_C\int_Cd^px \wedge d^py D(x-y)\right) \\ 
& =  \exp\left(-\frac{1}{2}g^2L^p \int_C d^px \int \frac{d^Dk}{(2\pi)^D}e^{ikx}\frac{1}{k^2}\right) \\ 
& =  \exp\left(-\frac{1}{2}g^2L^p \int \frac{d^{D-p}k}{(2\pi)^{D-p}} \frac{1}{k_\perp^2}\right),\ea \ee
where we have done the integral over $y$ by symmetry to add a multiplicative factor of $L^p$, 
and where $k_\perp$ represents the momentum components in the direction perpendicular to the submanifold $C$. Note how the integral over $C$ effectively reduces the problem to one involving the $D-p$ modes normal to the surface of $C$: the longitudinal modes along the brane are integrated out and do not affect the discussion of symmetry breaking. 

The final integral has a divergence at large $k$, which we will attempt to absorb by a multiplicative renormalization of the Wilson operators. If $\ln \langle W \rangle$ scales with $L$ as $-g^2L^p/a$ or slower (here $a$ is a short-distance cutoff), then $\langle W\rangle$ can be rendered non-zero by re-defining the Wilson loop by renormalizing it with the help of a local counterterm 
\be \label{renorm} W\mapsto W \exp\left(i c\int_C d^px\right),\ee 
where $c$ is some constant which will usually depend on the cutoff $a$. This multiplicative renormalization \eqref{renorm} allows us to renormalize the ``perimeter law'' $\ln \langle W_C\rangle \sim \exp(-g^2L^p/a)$ to a ``zero-law'' \cite{hastings2005quasiadiabatic}, where $|\langle W_C \rangle|=1$ for all $C$. Annoyingly, $c$ is non-universal and dependent on the details of the submanifold $C$ on which $W_C$ is defined. However, such shape dependence will be a $1/r$ effect (where $r$ is a radius of curvature) and so will not be important for us, since we are only interested in the IR phase structure and as such will always take $C$ to be large and smooth. 

More importantly, the integral \eqref{big_sheet_int} is logarithmically divergent at small momentum when $D-p = 2$. This IR divergence cannot be eliminated by a local UV counterterm, and tells us that $\langle W_C\rangle=0$ in the limit of infinite volume. This means that SSB, at least for Wilson operators that are supported on copies of $\rr^p$, is impossible if $p\geq D-2$. This is essentially because the integral over the Wilson brane $C$ dimensionally reduces the computation to one involving $D-p$ momentum modes, which are the ones that generate translations in the directions normal to $C$. 

In a similar way, we can also compute correlation functions of Wilson branes $\langle W_{C_0} W_{C_x}^*\rangle$, where $C_0$ and $C_x$ are two parallel copies of $\rr^p$ placed a distance $x$ away from one another. When $p=D-2$, the correlator has a form reminiscent of a Gaussian spin-wave two-point function, namely
\be \langle W_{C_0} W_{C_x}^*\rangle \sim \frac{1}{|x/a|^{g^2L^p}},\ee
which provides us with another way to see that the symmetry is unbroken. The form of the correlator also indicates that the theory is marginally confining: it is intermediate between a deconfined phase (where the correlator would approach a constant) and a normal confining phase (where the correlator would decay exponentially).

\subsubsection{Relation to confinement and Wilson operators on spheres} \label{confinement_stuff} 

We now briefly comment on the relation between spontaneous symmetry breaking for higher form symmetries and the problem of confinement in pure gauge theory. 
As we will see by examining whether or not Wilson operators are allowed to have nonzero expectation values, the $p$-form gauge theory is always in a confining phase when symmetry breaking is disallowed by dimensionality (i.e. when $p\geq D-2$), and in a deconfined phase in the symmetry broken state. This was noted in \cite{gaiotto2015generalized}; here we merely add some details. 

The diagnostic of confinement in $p$-form gauge theory at zero temperature is the scaling of the expectation value of a topologically trivial Wilson $p$-brane, which for simplicity we will take to be a $p$-sphere. If the expectation value vanishes as a ``perimeter law'' or slower, i.e. if in the large $R$ limit we have 
$\langle W_{S^p_R}\rangle 
\sim \exp\left(-g^2R^p/a\right),$
where $S^p_R$ is the $p$-sphere of radius $R$, then we say the theory is in the deconfined phase, while if the argument of the exponential diverges more rapidly than $R^p$ then we call the theory ``confining''. This criterion is because after multiplicatively renormalizing a perimeter law to a ``zero-law'' phase, the Wilson branes are tensionless and fluctuate on all scales (this phase can also be thought of as a $p$-form version of a string-net liquid \cite{levin2005string}).  
However, if $\ln \langle W_{S^p_R}\rangle$ diverges faster than a perimeter law, such a renormalization by a local counterterm is not possible, and so we can not render $\langle W_{S^p_R}\rangle$ finite in the large $R$ limit (and we will always be taking the $R\ra \infty$ limit, since we are only interested in Wilson operators that diagnose the IR physics). This is the confined phase, where the branes have tension. 

Of course, examining the behavior of the expectation values of Wilson operators is exactly what we've been doing with regards to examining SSB. However, while closely related, these two issues are slightly different in the way we approach them. This is because the operators $W_{S^p_R}$ are {\it uncharged} under the $p$-form symmetry: they are computed by integrating the gauge field $A$ over a topologically trivial submanifold of spacetime, and as such are invariant under the $p$-form symmetry action $A\mapsto A + \lambda$ with $\lambda$ a flat connection.
 A diagnostic of confinement that is more in-line with one of $p$-form symmetry breaking is a $p$-form generalization of the Polyakov loop operator, which, being computed by integrating around a non-contractible $p$-sphere, does transform nontrivially under a $p$-form symmetry. 

Now we will run through some example calculations in $D=2,3,4$. The general form of the answers can be obtained from dimensional analysis and anticipating the presence of logarithmic divergences in dimensionless integrals when $p=D-2$, but we still find it helpful to work out some examples in order to get a feel for how the $\langle W_{S^p_R}\rangle$ scale in different dimensions. 

{$\mathbf{D = 2}$}. Here the only interesting case is for $0$-forms, i.e. scalars. The story here is very familiar: to diagnose ``confinement'', which in this case is exactly the same as symmetry breaking, we compute $\langle W_{S^0_R}\rangle$ for a 0-form field $\phi$ in the limit of large $R$. Since $S^0_R$ is just two points separated by a distance of $2R$ and the propagator is $D(x-y) = -\frac{1}{2\pi} \ln|x-y|$, we get 
\be \ba \langle W_{S^0_R} \rangle 
& \sim \frac{1}{(2R)^{g^2/2\pi}}.\ea \ee
Since this vanishes as $R\ra \infty$, we see that $p=0$ theories ``confine'' in $D=2$: this is the regular CMW theorem. 

\medskip

{$\mathbf{D=3}$}. In three dimensions, we will need to look at both 0-form and 1-form theories.
For $p=0$ we have the scaling $\langle W_{S^0_R}\rangle \sim e^{g^2/R}$. This goes to a constant as $R\ra \infty$, and a ``deconfined'' (symmetry breaking) phase is allowed. 

For 1-form fields $A_\mu$, we let the Wilson loop be an $S^1$ of radius $R$. The 1-form propagator (in the Feynman gauge) is $D_{\mu\nu}(x-y) = \frac{\delta_{\mu\nu}}{4\pi |x-y|}$, and so we compute 
\be \ba \ln \langle W_{S^1_R} \rangle & = -\frac{g^2}{8\pi }\int_{S^1_R}\int_{S^1_R} dx^\mu \wedge dy^\nu \frac{\delta_{\mu\nu}}{|x-y|} \\
& = -\frac{Rg^2}{4}\int_{S^1_1} d\theta\, \frac{R\cos\theta}{\sqrt{2R^2 - 2R^2\cos\theta}}  \\ & 
= -\frac{g^2}{2}R\left(\ln(R/a)+2(\ln2-1)\right),\ea\ee
where we did the integral over $x$ by symmetry, added the $\cos\theta$ in the denominator because of the wedge product $dx\wedge dy$, and where $a$ is again a short-distance cutoff. The part which scales as $g^2R$ can be eliminated by adding a local counterterm, but the $\ln(R/a)$ piece diverges more strongly than a perimeter law and cannot be gotten rid of in such a manner. Thus, we conclude that the order parameter $\langle W_{S^1_R}\rangle$ cannot become non-zero in the large $R$ limit, and that the theory is logarithmically confining (it does not confine as an area law, i.e. $\langle W_{S^1_R}\rangle$ does not vanish as $e^{-g^2R^2/a}$, but in our definition it is still confining). 
The fact that the argument of the exponential is a logarithm suggests that while there is no confinement / deconfinement transition for regular gauge theories in $D=2+1$, a KT-like phase transition may be possible. 

{$\mathbf{D=4}$}.
For the scalar field, the propagator is now $\frac{1}{4\pi^2r^2}$, and so $\langle W_{S^0_R}\rangle \sim \exp(g^2/R^2)$; of course, a ``deconfining'' (symmetry-breaking) phase exists. 

We now consider a 1-form theory and a Wilson loop an $S^1$ of radius $R$, which we expect to have a perimeter law, consistent with having a symmetry-broken phase. We use $D_{\mu\nu}(x-y) = \delta_{\mu\nu}\frac{1}{4\pi^2|x-y|^2}$ to get 
\be\ba  \ln \langle W_{S^1_R} \rangle & = -\frac{g^2}{8\pi^2}\int_{S^1_R}\int_{S^1_R} dx^\mu \wedge dy^\nu \frac{\delta_{\mu\nu}}{|x-y|^2} 
\\ & = -\frac{g^2}{2\pi}\left(\frac{R}{a} -\frac{\pi}{2}\right),\ea\ee
which can renormalized to a ``zero-law'' by multiplicatively renormalizing the Wilson line. Thus the 1-form symmetry is allowed to be broken, and the photon is the Goldstone boson. 

Now for the 2-form theory, which for the same reason as the $p=1$ case for $D=3$ we expect to be logarithmically confining in $D=4$. 
The real-space 2-form propagator is 
\be D_{\mu_1\mu_2}^{\nu_1\nu_2}(x-y) = \frac{1}{4\pi^2}\frac{\delta_{\mu_1\mu_2}^{\nu_1\nu_2}}{|x-y|^2},\ee
which we obtain from \eqref{general_propagator}. 
We consider a Wilson surface which is an $S^2$ of radius $R$ and calculate 
\be \ba \ln \langle W_{S^2_R} \rangle & = -\frac{g^2}{8\pi^2}\int_{S^2_R}\int_{S^2_R} dx^{\mu_1} \wedge dx^{\mu_2} \wedge dy^{\nu_1} \wedge dy^{\nu_2} \frac{\delta_{\mu_1\mu_2}^{\nu_1\nu_2}}{|x-y|^2} \\ 
& = -\frac{g^2R^2}{4\pi}\int_{S^2_1} d\theta\, d\phi\, \frac{\cos\theta\sin\theta}{1-\cos\theta} 
\\ & = -g^2 R^2 \left(\ln(R/a) + (\ln2-1)\right).\ea\ee
This indeed diverges more strongly than a perimeter law, and the $\ln(R/a)$ piece cannot be renormalized away: thus, 2-form theories are always confining in $D=4$. 

In general spacetime dimension $D$, the pattern is clear. When $p<D-2$, we get a Wilson loop $W_C$ whose expectation value can be renormalized to a constant zero-law, but when $p=D-2$, we
get a Wilson loop whose expectation value on a $(D-2)$-sphere of radius $R$ scales as 
\be \langle W_{S^{D-2}_R} \rangle \sim \exp\left[-g^2R^{D-2}\left(\ln(R/a)+O(1)\right)\right]. \ee
This cannot be renormalized to a zero law, and thus its expectation value really does vanish in the large $R$ limit. Thus, while we find a scaling for the order parameter that would seem to allow for a higher-form type of KT transition for $p$-form theories in $p+2$ dimensions, no true deconfined phase exists in such cases. 

To summarize, we've seen that an analysis of confinement in $p$-form theories in various dimensions corroborates our thoughts about the relation between (de)confinement and higher form symmetry breaking: when $p\geq D-2$ only a confining phase exists, corresponding to the CMW-theorem statement that in such cases SSB is impossible.

\subsubsection{Computing expectation values with duality} \label{duality} 

We now turn a method of computing $\langle W_C\rangle$ and checking the higher CMW theorem using Abelian duality, which maps a $p$-form theory to a dual $(D-p-2)$-form theory. In particular, in the case of interest for the CMW theorem, namely $p=D-2$, the dual field is a scalar. This approach is especially useful for computing $\langle W_C\rangle$ when $C$ is nontrivial in $H_p(X;\zz)$ or when obtaining an explicit expression for the $p$-form propagator is difficult. 

Let us look at where a Wilson operator goes under duality (see also \cite{deligne1999quantum}). For a Wilson operator $W_C$, we have as before 
\be \langle W_C\rangle= \int \frac{\mcd A}{\vol(\mcg_{A,0})}\exp\left(-\int\left[\frac{1}{2g^2}F\wedge\star F-i A\wedge \wh C\right]\right),\ee
with $\mcg_{A,0}$ the group of trivial gauge transformations on $A$, which we will take to have Dirichlet boundary conditions. Next we add a field $K \in \Omega_N^{D-p-1}(X)$, which we will think of as a $(D-p-1)$-form gauge field with Neumann boundary conditions that has been gauge-fixed, satisfying the gauge constraint $d^\da K=0$. Ignoring overall constants, we get
\be \ba \langle W_C\rangle & = \int \frac{\mcd A\mcd K}{\vol(\mcg_{A,0})} \exp\left(-\int\left[\frac{g^2}{8\pi^2}K\wedge \star K+\frac{1}{2g^2}F\wedge\star F-i A\wedge \wh C\right]\right) \\
& =\int \frac{\mcd A\mcd K}{\vol(\mcg_{A,0})} \exp\left(-\int\left[\frac{g^2}{8\pi^2}K\wedge \star K+\frac{i}{2\pi}K\wedge F-i A\wedge \wh C\right]\right), \ea \ee
where we shifted $K\mapsto K+(-1)^{(p+1)(D-p-1)}\frac{2\pi i}{g^2}\star F$, which preserves $d^\da K=0$ since $dF=0$, and preserves the Neumann boundary conditions on $K$ since $\star F \in \Omega_N^{D-p-1}(X)$ as a consequence of $F|_{\p X}=0$. 

Now we introduce a $(D-p-2)$-form $\sca$, which will be dual to $A$. This is done at the expense of multiplying by $1/\vol(\mcg_{K,0})$, which is equivalent to ``un-doing'' the gauge-fixing by introducing gauge redundancy for $K$. $\sca$ can be thought of as the gauge part of $K$, and the un-doing of the gauge-fixing means making the replacement $K \mapsto K - d\sca$ in the action. If $D-p-2>0$, then $\sca$ comes with its own gauge redundancy, and so we must also divide by $\vol(\mcg_{\sca,0})$. Then
\be \ba \langle W_C \rangle & = \int \frac{\mcd A\mcd K\mcd\sca}{\vol(\mcg_{A,0})\vol(\mcg_{K,0})\vol(\mcg_{\sca,0})} \\ &\qquad \times \exp\left(-\int\left[\frac{g^2}{8\pi^2}(d\sca-K)\wedge \star (d\sca - K)+\frac{i}{2\pi}(K-d\sca)\wedge F-i A\wedge \wh C\right]\right). \ea \ee

We then integrate out $A$: the topologically nontrivial part of $F$ sets $K$ to be globally well-defined, and the globally-defined part of $A$ appears as $A\wedge (dK - (-1)^{(D-p)(p+1)}2\pi \wh C)$ and fixes the curvature of $K$. Thus we get
\be \ba \langle W_C \rangle
& = \int \frac{\mcd K \mcd \sca}{\vol(\mcg_{K,0})\vol(\mcg_{\sca,0})}   \delta(dK-(-1)^{(D-p)(p+1)}2\pi \wh C) \\ & \times \exp\left(-\frac{g^2}{8\pi^2}\int(d\sca-K)\wedge \star (d\sca-K)\right).
\ea \ee
In this path integral, we have a connection $K$ coupling to $\sca$, which is constrained to be such that it has a curvature $dK=2\pi\wh C$. 
Instead of writing the integrand as $(d\sca - K)\wedge\star(d\sca-K)$ however, we may also write it as $d\sca\wedge\star d\sca$ and do away with $K$ altogether, provided that we upgrade $\sca$ to a singular $(D-p-2)$ form which is well-defined away on $X\setminus C$, and is such that $\int_Nd\sca=2\pi$ for any closed $(D-p-1)$-manifold $N$ which links $C$. 

We should mention that if $X$ is closed and $C$ is a nontrivial class in $H_p(X;\zz)$, this procedure breaks down, as the holonomy condition on the dual field $\sca$ we obtain is inconsistent\footnote{For example, if $p=1,D=3$, then the dual field must vanish on a surface $B$ such that any loops linking $C$ intersect $B$ once. If $\p X=0$ we must have $\p B = C$, which is a contradiction if $C$ is nontrivial in $H^1(X;\zz)$ (see e.g. \cite{deligne1999quantum}).}. This is essentially the generalization of the fact that the field configuration of a single monopole on a compact, boundaryless manifold is inconsistent: there is nowhere for the Dirac string to go (on a non-closed manifold this is okay; we can have the string end on the boundary or go off to infinity). A consequence of this is that no operators that transform nontrivially under a $p$-form symmetry can have non-zero expectation values on closed manifolds (which we also observed from the Ward identity \eqref{ward_id}). 
Of course, if $X$ is closed we can still study the correlation functions of Wilson operators $\langle W_{C_1}\dots W_{C_n}\rangle$ provided that $C_1+\dots+C_n$ is trivial in $H_p(X;\zz)$ (charge neutrality). This won't be very important for usm, since we are interested in SSB and as such will be focusing on spacetimes which are not closed. 

The dual representation allows us to easily estimate the modulus $|\langle W_C\rangle |$. As an example, take $D=3,p=1$, and take space to be the tube $\Sigma=\rr \times S^1$. Consider a Wilson $W_C$, where $C$ is a straight line threading the $S^1$. In the dual theory, this maps onto a scalar field with a 1-form connection $K$ such that $dK = 2\pi \wh C$. 
As mentioned above, we can equivalently work with a scalar field $\phi$ without a connection which is well-defined on $X\setminus C$ and which winds by $2\pi$ around $C$. That is, we can write
\be \langle W_C\rangle = \int \mcd \phi \exp\left(-\frac{g^2}{8\pi^2}\int_{X\setminus C} d\phi \wedge \star d\phi\right),\ee
where we require that $\phi$ to have $2\pi$ monodromy along loops linking $C$. Note that a $2\pi$ winding is not trivial, because our theory has no dynamical matter: dynamical matter can screen integer-charge Wilson lines, but in its absence the Wilson lines remain unscreened.

Restricting our attention to a single plane orthogonal to $C$, we can estimate the minimal action in the plane by choosing the vortex solution
\be d\phi = d\theta  / r,\ee 
where $\hat\theta$ is the azimuthal unit vector in the plane and $r$ is the distance to the intersection of the plane with $C$. Then the minimal action of the theory restricted to the plane diverges logarithmically, and so the full minimal action is approximately 
\be S \sim \frac{g^2}{8\pi^2}|C|\ln(l/a),\ee
where $l$ and $a$ are long- and short-distance cutoffs and $|C|$ is the length of $C$. Since this diverges logarithmically as $l\ra \infty$, we conclude that $\langle W_C\rangle=0$ as a result of the fact that the action is infinite and unable to be multiplicatively renormalized to something finite. Therefore, we verify that the 1-form symmetry may never be spontaneously broken in this phase. 

This approach also allows us to compute correlation functions of Wilson lines. 
We consider the correlation function $\langle W_{C_0}W^*_{C_x}\rangle$, where $W_{C_0}$ and $W_{C_x}$ are two parallel Wilson lines placed a distance of $x$ from one another\footnote{Since $\langle W_{C_1}W^*_{C_2}\rangle$ is invariant under the $p$-form symmetry if $C_1$ and $C_2$ are in the same homology class, it can take nonzero values even on closed spacetimes.}. 

In the dual theory, the insertion of $W_{C_0}W^*_{C_x}$ tells us to consider a scalar field $\phi $ with additive monodromy $2\pi$ around $C_0$ and $-2\pi$ monodromy around $C_x$. 
We can find an approximate minimal-action solution by combining the $\star d\phi  = \pm d\theta / r$ solutions around both $C_0$ and $C_x$. The solution $d\phi = d\theta / r$ is a good approximation within a radius of $r\lesssim x/2$ from $C_0$, while $d\phi = - d\theta / r$ is an approximate solution within $r\lesssim x/2$ from $C_x$. At distances much larger than $x$ from the two Wilson lines, $\phi$ has trivial holonomy around loops linking both lines. Thus the minimal action should scale as $S \sim g^2 |C_0| \ln(x/a)$, and so the correlation function scales as 
\be \label{WW_correlation} \langle W_{C_0}W^*_{C_x}\rangle \sim \frac{1}{|x/a|^{g^2|C_0|}},\ee
which is characteristic of a marginally confining phase. 
This means that $\langle W_{C_0}W^*_{C_x}\rangle$ cannot be multiplicatively renormalized to something which does not vanish as $x\ra \infty$, and we again conclude that the 1-form symmetry cannot be spontaneously broken. 
 
We can also consider the same calculations, but one dimension up (viz. $D=4,p=1$). In this case, spontaneous symmetry breaking should be allowed. After duality, we have
\be \langle W_C\rangle = \int \frac{\mcd\sca}{\vol \mcg_{\sca,0}} \exp\left(-\frac{g^2}{8\pi^2}\int_{X\setminus C} d\sca \wedge \star d\sca\right),\ee
where this time $\sca$ is a 1-form field on $X\setminus C$ such that 
\be \int_{S^2}d\sca = 2\pi,\ee
for any $S^2$ that links $C$. 
We now examine the action on hypersurfaces normal to each point on the Wilson line. This time these cross-sections are volumes instead of surfaces, and the appropriate generalization of the vortex solution is to take
\be \star d\sca = \frac{1}{2r^2} dr \ee
on these volumes. Thus the minimal action is approximately 
\be S \sim \frac{g^2}{8\pi^2}|C|(1/a),\ee
which does not have an IR divergence and can be rendered finite through the usual multiplicative renormalization of the Wilson line. Since $\langle W_C\rangle$ can be renormalized to something non-zero, SSB is allowed. A similar computation gives the scaling $\langle W_{C_0} W^*_{C_x}\rangle \sim \exp(-|C|g^2/(8\pi x))$, which is non-zero as $x\ra \infty$ and again indicates that SSB is allowed. 
These arguments easily generalize to arbitrary $D$ and $p$, and allow us to check the higher CMW theorem in more general settings.

\subsection{Compact theories} \label{lattice}

In this section, we give a straightforward generalization of Polyakov's argument \cite{polyakov1977quark,polyakov1987gauge,metlitski2008wilson} for the confinement of ``compact'' $U(1)$ gauge theory in three dimensions to general $p$-form gauge theories in general dimensions, which establishes the higher CMW theorem for compact theories. 
We use ``compact'' in the sense that the theory is either defined on a lattice or possesses monopole operators which are included in the action. Since $d^\da \star F\neq0$ in this case, the magnetic symmetry (with conserved current $\star F$) is explicitly broken, but the electric symmetry (with conserved current $F$) remains intact. 

A heuristic picture is that if $p=D-2$ then $F\in \Omega^{D-1}(X)$, and so the Poincare dual $\wh F$ is a 1-dimensional string. Magnetic matter disorders the system by causing $dF\neq0$, and the Poincare dual statement is that $\p \wh F\neq0$, meaning that the $\wh F$ strings can end. The endpoints of the $\wh F$ strings are instantons, and since they are point-like, we expect that it will always favorable (in the free-energy sense) for instantons to proliferate. A Wilson operator which braids around an instanton transforms nontrivially and the presence of many instantons renders the photon massive, preventing SSB. 

To corroborate this picture, we start with the action \eqref{pform_action} and specify to the relevant case where $p=D-2$. 
If we have a lattice theory in mind, then integrals will be tacitly understood to represent sums and exterior calculus tacitly understood to involve cochains and coboundary operators. When formulated in terms of differential forms the manipulations are rather standard, so we will be brief. 

We start by Hodge decomposing $F$ as 
\be F = dA + 2\pi(d^\da \beta + \omega).\ee
Here $A\in \Omega^{D-2}(X)$ is a globally well-defined form, $\beta \in \Omega_\zz^D(X)$ is a $D$-form with integer periods, and the harmonic component $\omega \in \Omega^{D-1}_\zz(X)$ is a $(D-1)$-form which will play a passive role in what follows.
Since $d^\da \beta + \omega$ has integer periods, $F$ is properly quantized. 

The action is 
\be S = \frac{1}{2g^2} \left( ||dA||^2 + 4\pi^2 (|| d^\da \beta ||^2 + ||\omega||^2)\right),\ee 
where $||a||^2 = \langle a,a\rangle$ with the usual inner product $\langle a,b\rangle = \int a\wedge \star b$. To write $S$ in this way, we have used the fact that the terms in the Hodge decomposition of $F$ are mutually orthogonal with respect to $\langle\; ,\; \rangle$. 

We now run Abelian duality on the $(D-2)$-form gauge field $A$ in essentially the same way as in \ref{duality}. This is done by doing a functional Fourier transform
\be \frac{1}{2g^2} ||dA||^2 \leftrightarrow \frac{1}{2g^2}||dA - K||^2 + \frac{i}{2\pi}\int d\phi \wedge K,\ee
where on the RHS, $K$ is a $(D-1)$-form gauge field and $\phi$ is a scalar that will end up as the dual to $A$ (more generally if $A$ were a $p$-form, $\phi$ would be a $(D-p-2)$-form). 

Now we use $K$ to gauge fix $A$ to zero (thus working in ``unitary gauge'')
and then make the shift $K\mapsto K - \frac{ig^2}{2\pi}\star d\phi,$ which kills the $d\phi \wedge K$ term and lets us integrate out $K$. 
Doing a further shift $\phi \mapsto \phi + \frac{4\pi^2}{g^2}\star \beta$ 
cancels the $||d^\da \beta||^2$ term, and leaves us with 
\be S = -\frac{g^2}{2(2\pi)^2}||d\phi||^2 + \frac{(2\pi)^2}{2g^2}||\omega||^2 - \int \phi \wedge \star \Delta \beta,\ee
where we have used $dd^\da \beta=\Delta\beta$ since $d\beta=0$ by virtue of $\beta$ being top-dimensional. 

We now introduce a $\zz$-valued $D$-form $q$ which keeps track of the location of spacetime instantons of the gauge field. We write $q = \sum_{x_i} q_i \star \delta(x-x_i)$, where the $x_i$ are the locations of points around which the flux of $F$ is $2\pi q_i$, with $q_i\in \zz$. That is, we write 
\be q_i = \int_{S^{D-1}_i} \frac{F}{2\pi},\ee
where $S^{D-1}_i$ is a small sphere surrounding the point $x_i$. Thus rather than defining the theory by excising the $x_i$ points from $X$ and having $dF=0$ globally, we will let $F$ be defined at $x_i$ and allow $dF\neq0$. Recapitulating, we have
\be q = dF = \Delta \beta,\ee
since $d^2A=d\omega =0$. 
Therefore, we obtain the final form of the action for the dual scalar:
\be S = -\frac{g^2}{2(2\pi)^2}||d\phi||^2 + \frac{(2\pi)^2}{2g^2}||d \omega||^2 + \int\phi \wedge q.\ee

To diagnose symmetry breaking we examine the expectation value of the monopole current 1-form $\star F$, which maps simply under duality. Without the instantons $\star F$ is massless, but with the instantons it acquires a mass. The topological charge of the instantons is measured by $d^\da (\star F)$, the vanishing of which is obstructed by the non-flatness of the dual potential $\wt A$, which we can see by Hodge decomposing $\star F$. To find the correlations functions of $\star F$, we compute the generating functional  
\be Z[\xi] = \left\langle \exp \left( i \int \star F \wedge \star \xi \right)\right\rangle =\left\langle \exp\left(i\int A\wedge d\xi\right)\right\rangle\ee
for a 1-form probe field $\xi$. Thus the generating functional for $\star F$ is a Wilson loop $W_C$ with $C=\wh{d\xi}$.

Tracing this through abelian duality, we find that the expectation value is given by 
\be\ba Z[\xi] = \int \mcd \phi \mcd q \mcd \omega& \exp\left(-S[\phi,\omega,q]+\frac{g^2}{4\pi}\int (2d\phi \wedge \star \xi +2\pi \xi\wedge \star \xi) \right),\ea\ee
where $S[\phi,\omega,q]$ is the action with $\xi =0$. 

We can then compute correlation functions of $\star F$ by computing $(\delta Z[\xi] / \delta \xi)|_{\xi =0}$:
\be \ba 
\langle (\star F)_\mu(k)(\star F)_\nu(-k) \rangle & = \frac{g^2}{2\pi}\left(2\pi\delta_{\mu\nu} - \frac{g^2}{2\pi}k_\mu k_\nu \langle\phi(k)\phi(-k)\rangle|_{\xi =0}\right).\ea\ee
If $dF=0$ so that $q=0$, $\phi$ would be a regular massless free scalar, and we would obtain a massless vector propagator for $\star F$. With $dF\neq0$, the correlation function of $\phi$ changes in such a way to make the $\star F$ propagator massive (see also \cite{polyakov1987gauge}), since $\phi$ picks up a mass from the $\phi \wedge q$ term. If we make the usual approximation where only charge $q=\pm1$ instantons contribute, we proceed as in the $D=3,p=1$ case to obtain a propagator with a mass $m^2 = 2\pi e^{-c/g^2}/a^Dg^2$, where $a$ is a characteristic short-distance instanton radius and $c$ is a constant relating to the instanton self-energy.

Since we have a massive theory, we expect that no spontaneous symmetry breaking is allowed. 
We can check this by estimating the expectation values of Wilson loops. The expectation value of a Wilson operator $W_C$ is $Z[\xi]$, with $\xi$ defined by $d\xi = \wh C$. 
From our expression for $Z[\xi]$, we see that $Z[\xi]$ can be obtained from the source-free partition function by making the replacement $d\phi \mapsto d\phi - 2\pi  \xi$.
So just as before, $\xi$ becomes a connection for $\phi$ that forces $\phi$ to wind by $2\pi$ when taken around $C$. 

Now we follow the usual argument by considering the case when $C$ is very large \cite{polyakov1987gauge}. The minimal action for $\phi$ with an insertion of the Wilson loop will have to wind by $2\pi$ on loops linking $C$. 
Since $\phi$ is massive, this winding will cost a large amount of action, and the minimal action will scale as the area of the minimal surface bounded by the Wilson loop, which is where the winding of $\phi$ will be localized. This leads to a vanishing expectation value for $W_C$ in the limit of large $C$, and implies that the $p$-form symmetry can never be spontaneously broken. Note that the underlying mechanism responsible for the preservation of symmetry in this case is rather different than in the theory with no monopole operators: in this case the photon becomes massive and the absence of massless bosons rules out spontaneous symmetry breaking, whereas in the theory without monopoles the photon remains massless but does not correspond to a Goldstone mode, because strong fluctuations prevent the existence of an ordered state.

\subsection{Discrete theories} \label{discrete}

We now briefly make a few comments on the case where the $p$-form symmetry is discrete. The analogue of the CMW theorem for the discrete case is the obvious modification of the CMW for the continuous case by a change in one in the critical dimension, namely that discrete $p$-form symmetries in $D$ spacetime dimensions cannot be broken at any finite temperature if $p\geq D-1$ \cite{gaiotto2015generalized}. Indeed, higher form discrete symmetries must be able to be broken for $p=D-2$, since topological phases in $2+1$ dimensions provide us with examples of such symmetry-breaking phases. One consequence of the result is that it allows us with an alternate derivation of the well-known fact that (bosonic) topological order cannot exist in one dimension \cite{chen2011classification}. 

One can argue this from a tiny generalization of the free energy arguments given for the $p=0$ case. For regular 0-form symmetries, the objects that disorder the system are $(D-1)$-dimensional domain walls. When $D=1$ the domain walls become zero-dimensional, and it becomes entropically favorable to proliferate them (assuming that they have finite energy). For 1-form theories, the domain walls are $(D-2)$-dimensional (in $D=3$ these domain walls are the Wilson lines of topological field theories or the strings in string-net models), and so when $D=2$ the domain walls are zero-dimensional objects, and at any finite temperature they proliferate, destroying the order. 

For general $p$ the domain walls are $D-p-1$ dimensional objects, and so when $p=D-1$ they proliferate and symmetry breaking is disallowed. 
Of course, when $p<D-1$, we can have phase transitions. For example, if $p=D-2$, then the domain walls are one dimensional. Assuming that the energy cost of a disordering operator supported on a string of length $L$ is proportional to $L$, the free energy cost of creating such a string is $F\sim \alpha L - T\ln(\#)$, where $\alpha$ is a constant and $\#$ is the number of strings of length $L$. On a lattice the latter scales as $c^L$ for some constant $c$, and so $F\sim L(\alpha - T\ln c)$, which changes sign at a finite value of $T$.

\section{Discussion} \label{discussion} 

In this paper, we have offered some comments on how the behavior of theories with spontaneously broken conventional symmetries generalizes to the case where the symmetries are higher $p$-form symmetries. It is natural to wonder about how other standard results extend to the general $p$ case. For example, for continuous theories the scaling of Wilson operators (namely the logarithmically confining properties of $p=D-2$ theories) has given us hints of the presence of KT-like phase transitions when $p=D-2$. It would be interesting to think about these types of phase transitions in more detail. 

The potential use of higher symmetries as a tool for describing asymptotic symmetries in Abelian gauge theory is also interesting. Currently this connection is still rather superficial, and it is natural to wonder if other things relating to asymptotic symmetries, like the memory effect and soft theorems, can be interpreted from a higher symmetry standpoint. It could also be illuminating to run through some more careful examples in other spaces like AdS space, where boundary condition issues are likely more straightforward to deal with.

Finally, we have had comparatively little to say about the $(D-p-2)$-form magnetic symmetries. It would be nice to be able to better understand the interplay between the electric and magnetic symmetries and the consequences of the mixed anomaly between them \cite{gaiotto2015generalized}, as well as to be able to formulate boundary conditions
that allow for a more unified treatment of both symmetries.

\acknowledgments
I am in debt to Sa\v so Grozdanov, Daniel Harlow, Robert Jones, Hong Liu, Max Metlitski, Jake McNamara, Sabrina Pasterski, and Ryan Thorngren for discussions and for patiently answering questions. I am supported by the Fannie and John Hertz Foundation and the NDSEG fellowship. 

\appendix
\section{Conventions for exterior calculus} \label{forms}

In this appendix, we quickly summarize our conventions for exterior calculus. 
A $p$-form $A\in\Omega^p(X)$ has the usual factor of $1/p!$ when written out component-wise:
\be A = \frac{1}{p!}A_{\mu_1\dots\mu_p}dx^{\mu_1}\wedge\dots\wedge dx^{\mu_p}.\ee
The Hodge star of a form $A\in \Omega^p(X)$ is 
\be \star A = \frac{\sqrt{|g|}}{p!(D-p)!}A_{\mu_1\dots\mu_p}\varepsilon^{\mu_1\dots\mu_p}_{\nu_{p+1}\dots\nu_D}dx^{\nu_{p+1}}\wedge\dots\wedge dx^{\mu_D},\ee
where $D=\dim X$. When acting on $p$-forms, it satisfies 
\be \star\star = (-1)^{p(D-p)+q},\ee
where $q=1$ $(q=0)$ for Lorentzian (Euclidean) signature. 
The volume form is 
\be \vol =\star 1= \frac{\sqrt{|g|}}{D!}\varepsilon_{\mu_1\dots\mu_D}dx^{\mu_1}\wedge\dots\wedge dx^{\mu_D} = \sqrt{|g|}dx^1 \wedge\dots\wedge dx^D.\ee
Since the Hodge star maps $p$-forms to $(D-p)$-forms, it gives us an inner product on $\Omega^D(X)$, given by integration: $\langle A, B\rangle = \int_X A \wedge \star B$. The inner product is symmetric:
\be \int A\wedge \star B = \int B\wedge \star A,\qquad \int A\wedge B = \int \star A \wedge \star B.\ee
The adjoint of $d$ is defined in the usual way, namely by $\langle A,dB\rangle = \langle d^\da A,B\rangle$. On manifolds with $\p X \neq 0$, $d$ and $d^\da$ are only adjoint when acting on forms satisfying either Dirichlet or Neumann boundary conditions. When $d^\da$ acts on $p$-forms, it has the representation $d^\da = (-1)^{Dp + D +1}\star d \star$ in Euclidean signature, 
while it flips sign for Lorentzian signature. The Hodge Laplacian is $\Delta= (d^\da + d)^2$, which commutes with both $d$ and $d^\da$. 

Poincare duality provides a way to associate $k$-forms with codimension $k$ submanifolds (see e.g. \cite{bott2013differential,bredon2013topology}). In an ambient $D$-dimensional manifold $X$, if $A$ is any $D-p$ form and $B$ any $p$-form, then 
\be \int_X A \wedge B = \int_{\wh B \subset X} A,\ee
where $\wh B$ is the Poincare dual of $B$ (technically, the compact Poincare dual). 
Conversely, if $N\subset X$ is a $D-p$ dimensional submanifold of $X$, then $\int_N A = \int_X A \wedge \wh N$, where $\wh N$ is a $p$-form (equal to the Thom class of the normal bundle of $N$ in $X$). By applying Poincare duality to a wedge product $A\wedge B$, we have
\be \int_X A \wedge B = \int_{\wh A\cap \wh B} 1,\ee
telling us that the wedge product of two forms is Poincare dual to the intersection product, and so $\int_X A\wedge B$ simply represents the intersection of $\wh A$ and $\wh B$. The intersection $\cap$ is signed, so that for a $m$-manifold $M$ and an $n$-manifold $N$, both embedded inside a $D$-manifold $X$, we have $M \cap N = (-1)^{(D-m)(D-n)}N\cap M$. We can also write the integral as
\be \int_X A\wedge B = \sum_{p\in(\wh A \cap \wh B) \subset X} {\rm sgn}(p),\ee 
where ${\rm sgn}(p)=\pm1$ according to whether the orientation of $T_p\wh A\oplus T_p\wh B$ agrees or disagrees with that of $T_pX$. 

Poincare duality sets up an isomorphism 
\be H^p(X) \cong H_{D-p}(X),\ee 
where the coefficients can be in any unital ring. 
When $\p X$ is non-zero, this changes slightly: we instead have 
\be H^p(X,\p X) \cong H_{D-p}(X), \qquad H_p(X,\p X) \cong H^{D-p}(X).\ee 
Here the relative (co)homology groups are such that elements in $H^p(X,\p X)$ are $p$-forms that vanish on $\p X$, while elements in $H_p(X,\p X)$ are $p$-submanifolds whose boundaries are contained within $\p X$ (i.e. submanifolds which are closed modulo $\p X$). This isomorphism also holds on the chain-cochain level, so that e.g. $Z^p(X,\p X)\cong Z_{D-p}(X)$. 

The Poincare dual of a given submanifold can always be made to have support only within a tubular neighborhood of the submanifold. 
For example, the poincare dual of a point $p\in X$ is (a smoothened version of) the delta function $\delta(x-p) dx^1\wedge \dots \wedge dx^D$, while the Poincare dual of $X$ is the 0-form constant function $1$. In $\rr^2$, the Poincare dual of the $y$-axis is $\rho(x)dx$, where $\rho(x)$ is an arbitrarily-narrow bump function centered on $x=0$. 
If we don't care about smoothness, we can use the integral expression 
\be \wh C_{\mu_{p+1}\dots\mu_D}(x) = \frac{1}{p!}\int_C \varepsilon_{\mu_1\dots\mu_p\mu_{p+1}\dots\mu_{D}}\delta(x-y)dy^{\mu_{1}} \wedge \dots \wedge dy^{\mu_{p}} \ee
for the components of the Poincare dual of a $p$-dimensional submanifold $C$. 

\bibliographystyle{jhep}

\bibliography{SSB_and_higher_symmetries_bib}

\end{document}